# Non-Bloch Theory for Spatiotemporal Photonic Crystals Assisted by Continuum Effective Medium

Haozhi Ding and Kun Ding†

*Department of Physics, State Key Laboratory of Surface Physics, and Key Laboratory of Micro and Nano Photonic Structures (Ministry of Education), Fudan University, Shanghai 200438, China*

† Corresponding E-mail: kunding@fudan.edu.cn

**Abstract**

As one indispensable type of nonreciprocal mechanism, a system with temporal modulations is intrinsically open in the physical sense and inevitably non-Hermitian, but the space and time degrees of freedom are nonseparable in a large variety of circumstances, which restrains the non-Bloch band theory to apply. Here, we investigate the spatially photonic crystals (PhCs) composed of spatiotemporal modulation materials (STMs) and homogeneous media, dubbed as the STM-PhC, wherein the spatial and temporal modulations are deliberately designed to be correlated. To bypass the difficulty of the spatiotemporal correlation, we first employ the effective medium theory to account for the dispersion of fundamental bands under the influence of Floquet sidebands. Based on the dynamical degeneracy splitting viewpoint and continuum generalized Brillouin zone condition, we then analytically give the criteria for the existence of the non-Hermitian skin effect in the STM. Assisted by developing a numerical method that embeds the plane wave expansion in the transfer matrix, we establish the non-Bloch band theory for the low-frequency Floquet bands in the STM-PhCs, in which the central is the identification of the generalized Brillouin zone. We finally delve into the topological properties, including non-Bloch Zak phases and delocalization of topologically edge states. Our work validates that effective medium assists the non-Bloch band theory applied to the STM-PhCs, which delivers a prescription to broaden the horizons of non-Bloch theory.

## Section I. Introduction

Recent growing efforts have been devoted to a myriad of physically open systems, ranging from condensed matter to classical waves, thus leading to non-Hermitian physics [1-6]. The allure of non-Hermitian physics lies in its complex spectrum, although the non-Hermitian systems with pseudo-Hermiticity can still possess real spectra [7-12]. The complexification immediately gives rise to two intriguing spectral features unique to non-Hermitian systems, known as exceptional degeneracy and point gaps [3,4,13-16]. The former handles an abundance of geometry formed by exceptional points and corresponding properties, including higher-order exceptional lines [17,18], eigenvalue braidings [19,20], non-Abelian conservation rule [21], and so on [3,4,13,15,22-25]. The latter indicates the non-zero eigenvalue winding numbers and further implies the wavefunction localization to the system boundary under open boundary conditions (OBCs), dubbed as the non-Hermitian skin effect (NHSE) [26-30]. The emergence of NHSE has led to the failure of the bulk-boundary correspondence based on the Hermitian Bloch band theory [26-28]. For restoring it, one prevailing approach is employing the generalized Brillouin zone (GBZ) to build up a comprehensive non-Bloch band theory, which works excellently in one-dimensional (1D) systems [26,31-33]. In higher dimensions, the non-Bloch band theory and the GBZs still have the same role as in 1D, although how to acquire the OBC spectra and wavefunctions from the non-Bloch band theory for a particular class of higher-dimensional systems is still under exploration [34-40].

Aiming to investigate the non-Hermitian photonic crystals (PhCs), the fact that the non-Bloch band theory is concomitant with the NHSE requires nonreciprocal electromagnetic materials [41] since the NHSE in lattice models has been achieved mainly by using nonreciprocal hoppings [26-28]. Besides using external fields or nonlinearity, time modulation is another way to break reciprocity [41,42], and by using it, a plethora of schemes have then been proposed to realize the phenomena unique in nonreciprocity, such as temporal double-slit interference [43], Fresnel drag effect [44], axion responses [45], time crystals [46-48], and so on [49-54]. As a spontaneously nonreciprocal material, the spatiotemporal modulation materials (STMs) are excellent candidates for the component in non-Hermitian PhCs, wherein the existence of exceptional points has been revealed [55-58]. The occurrence of NHSE is then seemingly apparent, and so is the non-Bloch band theory, but establishing it in the STMs is not straightforward because of the inherently $(d+1)$-dimensional problem herein ($d$ is spatial dimensionality). Moreover, most studies on spatiotemporal modulated systems nowadays deal with cases where space and time are separable or possess a uniform modulation speed in the space-time domain, certainly not fully leveraging the role of temporal degrees of freedom in non-Hermitian physics [44,45,54-58].

By targeting the establishment of non-Bloch band theory for the $(1+1)$-dimensional PhCs, we spatially stack the STM possessing a traveling-wave modulation in its permittivity and permeability with other homogeneous media, constituting the so-called STM-PhCs. Firstly, we in Sec. II deploy the effective medium theory (EMT) to represent the fundamental Floquet band of the STM influenced by the Floquet sidebands, which are benchmarked with the plane wave expansion (PWE). By further using the dynamical degeneracy splitting (DDS) viewpoint, we analytically work out the criterion for the existence of NHSE in STM without evoking the OBCs (Sec. II). To be precise, we then dwell on the OBC spectra and electromagnetic fields of the STM under different boundary conditions (BCs), and the continuum GBZ condition for the STM has validated both the criterion from DDS and the OBC results (Sec. III). With the recipe for STM in hand and by generalizing the transfer matrix method (TMM) through embedding the PWE, we establish the non-Bloch band theory for the STM-PhCs by identifying the GBZ and the OBC spectra and electromagnetic fields (Sec. IV). To validate the established non-Bloch band theory, we explore the non-Hermitian topological behaviors from non-Bloch bulk-boundary correspondence to delocalization of topological edge states (TESs) in Sec. V. The discussions and conclusions are drawn in Sec. VI.

**Section II. Manifestation of skin modes from effective medium theory**

The non-Bloch band theory requires knowing the map $f$ and its inverse $f^{-1}$ from the complex frequency (energy) domain $\mathbb{C}$ to the complex wavenumber domain $\mathbb{C}^d$ [26,31,32,35], which is demanding to acquire fully in the STM-PhCs because of the interplay between the wavenumber and Floquet quasi-energy. The EMT, a method only valid in the long-wavelength limit, establishes the map $f$ analytically, thereby mitigating the complexity of using non-Bloch band theory [59,60]. Aiming to utilize such analyticity and convenience, we establish the EMT for the STM with non-Hermitian modulations in this section, which embed the influence of Floquet sidebands in the zeroth band analytically as the effective parameters. This allows the analysis of NHSE from the wave propagation point of view firstly and spontaneously links to non-Bloch band theory for the STM-PhCs, which will be established later. For simplicity, the permittivity and permeability under investigation are of a traveling-wave manner as

$$\boldsymbol{D}(x,t)=\varepsilon_0\varepsilon(x,t)\boldsymbol{E}(x,t),\qquad \boldsymbol{B}(x,t)=\mu_0\mu(x,t)\boldsymbol{H}(x,t),\tag{1}$$

$$\varepsilon(x,t) = \varepsilon_r[1 + 2\alpha_\varepsilon \cos(gx - \Omega t + \phi_\varepsilon)], \tag{2}$$

$$\mu(x,t) = \mu_r[1 + 2\alpha_\mu \cos(gx - \Omega t + \phi_\mu)], \tag{3}$$

where $\varepsilon_r$ ($\mu_r$) represents the background permittivity (permeability), $\alpha_\varepsilon$ ($\alpha_\mu$) represents the complex modulation of permittivity (permeability), $\phi_\varepsilon$ ($\phi_\mu$) denotes the initial phase of the modulation in permittivity (permeability), and $g$ ($\Omega$) is the spatial (temporal) frequency. Figure 1(a) depicts a typical spatiotemporal modulation of material parameters and the corresponding Floquet band structure (BS) calculated by the PWE for $E_z$ polarization is shown in Fig. 1(b). We generalize the PWE algorithm established for the STM with $\alpha_\varepsilon$ and $\alpha_\mu$ being real numbers [44] to the PhCs with STM components included and $\alpha_\varepsilon$ and $\alpha_\mu$ being complex (see Appendix A for details). The leading three sets of bands, namely the fundamental order and two sidebands of order $\pm 1$, have been displayed in Fig. 1(b), wherein several frequency gaps are from the interaction between the fundamental bands and sidebands. If we focus on the long-wavelength limit that EMT works, the bands highlighted in the red dashed box, which connect to two asymmetric band gaps stemming from sidebands, can possess various features, including slopes and patterns, by tuning $\alpha_\varepsilon$ and $\alpha_\mu$. When the modulations $\alpha_\varepsilon$ and $\alpha_\mu$ are real numbers, the energy bands lean to a particular $k$ direction, indicating the occurrence of the Fresnel drag effect [44]. Figure 1(c) intentionally shows the long-wavelength bands with $\alpha_\varepsilon = \alpha_\mu = 0.2$ (open circle markers) and $\alpha_\varepsilon = \alpha_\mu = 0$ (red dashed lines). The conical dispersion leans to the positive $k$, meaning spectral nonreciprocity $\omega^P(+k) \neq \omega^P(-k)$ and $k_+(\omega^P) \neq k_-(\omega^P)$, where the superscript $P$ denotes the spectra under periodic boundary conditions (PBCs) and the subscript $\pm$ represents the wave propagation direction. Viewing from the equal frequency contour (EFC), we can interpret such an inclined dispersion as experiencing a gauge potential [top panel in Fig. 1(d)]. From the wave propagation point of view, the two waves $k_+(\omega^P)$ and $k_-(\omega^P)$ still interfere to form the solutions under OBC, resulting in the nonreciprocal response of the STM [bottom panel in Fig. 1(d)]. However, both viewpoints are limited to the real axis of $\omega$ and $k$ and do not imply the NHSE, thus begging for a perception extended into the complex plane.

When the modulations $\alpha_\varepsilon$ and $\alpha_\mu$ become complex numbers, the frequency ω and wavenumber $k$ inevitably shall enter the complex plane. Without evoking non-Bloch band theory, $\omega$ and $k$ cannot be simultaneously treated as independent complex variables because the method in Bloch band theory is to vanish the determination of one matrix equation. Hence, there are two alternative approaches: fixing $k$ or $\omega$ real. Setting $k$ as real numbers (within the BZ) leads to complex ω, denoted as the PBC or Bloch BS. Setting $\omega$ as real numbers leads to complex $k$, sometimes dubbed the complex-$k$ BS [61,62], which is a crucial ingredient in wave propagation. In the Hermitian scenario, these two complement each other and reflect different aspects of a physical system, and so does the STM with complex $\alpha_\varepsilon$ and $\alpha_\mu$.

Figure 1(e) shows the PBC BS when $\alpha_\varepsilon = 0.2i$ and $\alpha_\mu = 0.2$ calculated by PWE (open circle markers). The real and imaginary parts of $\omega^P$ are displayed by the left (in blue) and right (in red) y-axes. The PBC spectra are clearly nonreciprocity $\omega^P(k_+ = +k) \neq \omega^P(k_- = -k)$, which is the same as Fig. 1(c), but have an intriguing difference from $\mathrm{Im}(\omega^P)$ [37]. If we still inspect the EFC, which now means the equal $\mathrm{Re}(\omega^P)$ contour, besides the asymmetry in $k$, what is more profound is the separated $\mathrm{Im}(\omega^P)$, as shown in the top panel of Fig. 1(f). The two PBC states on the EFC have disparate lifetimes $[\mathrm{Im}(\omega^P)^{-1}]$, as depicted by the colors of the dots, hinting that these two Bloch waves under the OBC cannot interfere to form a bulk state, and the NHSE can then occur. The bottom panel of Fig. 1(f) schematically shows such two waves with the $k_+$ $(k_-)$ states having negative (positive) $\mathrm{Im}(\omega^P)$, thus leading to the tendency of left-going skin modes. This DDS perception successfully predicts the existence of NHSE in higher-dimensional spatial crystals and also works well in the spatiotemporal crystals here [37].

To see whether the complex-$k$ BS also implies the tendency of NHSE, we now investigate Fig. 1(g), showing the complex-$k$ BS with the same parameters as Fig. 1(e). The real and imaginary parts of $k$ are displayed by the bottom (in blue) and upper (in red) x-axes. The branches propagating towards $+x$ $(-x)$ are labeled by open squares (open circles), and the $\mathrm{Im}(k)$ of both branches are non-zero and nearly identical in the long-wavelength limit. If we plot the values of $k$ at a fixed ω in the complex wavenumber plane, the STM can be effectively treated as experiencing an imaginary gauge potential instead of a real one [top panel in Fig. 1(h)] [63]. This further indicates that all the waves, whatever the values of $\mathrm{Re}(k)$ are, tend to concentrate towards the $-x$ direction, also implying the left-going skin modes [bottom panel in Fig. 1(h)]. Note that for the $\omega$ within the band gap of a Hermitian system, the values of $k$ are also complex but symmetric about the origin, and no gauge potential exists [64].

The above two perceptions qualitatively imply the same tendency of NHSE without introducing the OBC, and we now demonstrate by EMT that the DDS-based correspondence yields the criterion for determining the presence of NHSE, which will be further proved in Sec. III by the non-Bloch theory for the continuum system. In the long-wavelength region, the STM defined by Eqs. (1-3) can be modeled as a homogeneous bianisotropic material (see Appendix B for details) [44,65]

$$\begin{pmatrix} D_x \\ D_y \\ D_z \\ B_x \\ B_y \\ B_z \end{pmatrix} = \begin{pmatrix} \varepsilon_0\varepsilon_{\text{eff},x} & 0 & 0 & 0 & 0 & 0 \\ 0 & \varepsilon_0\varepsilon_{\text{eff},y} & 0 & 0 & 0 & +c^{-1}\xi_{\text{eff}} \\ 0 & 0 & \varepsilon_0\varepsilon_{\text{eff},z} & 0 & -c^{-1}\xi_{\text{eff}} & 0 \\ 0 & 0 & 0 & \mu_0\mu_{\text{eff},x} & 0 & 0 \\ 0 & 0 & -c^{-1}\xi_{\text{eff}} & 0 & \mu_0\mu_{\text{eff},y} & 0 \\ 0 & +c^{-1}\xi_{\text{eff}} & 0 & 0 & 0 & \mu_0\mu_{\text{eff},z} \end{pmatrix} \begin{pmatrix} E_x \\ E_y \\ E_z \\ H_x \\ H_y \\ H_z \end{pmatrix}, \quad (4)$$

where $c = 1/\sqrt{\varepsilon_0\mu_0}$, $v = 1/\sqrt{\varepsilon_0\mu_0\varepsilon_r\mu_r}$, and the effective medium parameters are

$$\varepsilon_{\text{eff},x} = \varepsilon_r, \qquad \mu_{\text{eff},x} = \mu_r, \quad (5)$$

$$\varepsilon_{\text{eff},y} = \varepsilon_{\text{eff},z} = \varepsilon_r\left(1 + \alpha_\varepsilon^2 \frac{2\Omega^2}{v^2g^2 - \Omega^2}\right), \mu_{\text{eff},y} = \mu_{\text{eff},z} = \mu_r\left(1 + \alpha_\mu^2 \frac{2\Omega^2}{v^2g^2 - \Omega^2}\right), \quad (6)$$

$$\xi_{\text{eff}} = \alpha_\varepsilon\alpha_\mu \frac{2cg\Omega}{v^2g^2 - \Omega^2}\cos(\phi_\varepsilon - \phi_\mu). \quad (7)$$

The matching dispersion relation and eigenmodes for $E_z$ polarization are

$$k_\pm = \frac{\omega}{c}\xi_{\text{eff}} \pm \frac{\omega}{c}n_{\text{eff}}, \qquad \psi_\pm = \begin{pmatrix} 1 \\ \mp\frac{1}{Z_{\text{eff}}} \end{pmatrix}, \quad (8)$$

where the subscript $\pm$ denotes the waves propagating along $+x$ and $-x$ directions, $n_{\text{eff}} = \sqrt{\varepsilon_{\text{eff},z}\mu_{\text{eff},y}}$, $Z_{\text{eff}} = \sqrt{\mu_{\text{eff},y}/\varepsilon_{\text{eff},z}}$, and $\psi = (\sqrt{\varepsilon_0}E_z, \sqrt{\mu_0}H_y)^T$. The dispersions by Eq. (8) are displayed by solid lines in Figs. 1(c), 1(e), and 1(g), and excellent agreement with the PWE is seen when $\text{Re}\,\omega/\Omega < 0.15$. When $\text{Re}\,\omega/\Omega > 0.15$, the comparison between EMT and PWE is still qualitatively correct, so we derive the analytical criterion for the occurrence of NHSE in the STM from the EMT perspective by using the correspondence between DDS and NHSE [37].

Firstly, it is necessary to determine the wave numbers on the EFC from the PBC spectra. Within the EMT here, there are only two possible wave numbers $k_\pm$ for a fixed $\text{Re}(\omega^P)$. By requiring the same $\text{Re}(\omega^P)$ in Eq. (8), we obtain the ratio between $k_+$ and $k_-$ on the EFC

$$\frac{k_+}{k_-} = \frac{(\xi'_{\text{eff}} - n'_{\text{eff}})[(\xi'_{\text{eff}} + n'_{\text{eff}})^2 - (\xi''_{\text{eff}} + n''_{\text{eff}})^2]}{(\xi'_{\text{eff}} + n'_{\text{eff}})\left[(\xi'_{\text{eff}} - n'_{\text{eff}})^2 - (\xi''_{\text{eff}} - n''_{\text{eff}})^2\right]}, \quad (9)$$

where the prime and double prime superscripts denote the real and imaginary parts, respectively. The above ratio is $-1$ when $\xi_{\text{eff}} = 0$, indicating spectral reciprocity, and hence, the STM is generally spectral nonreciprocal because it breaks Lorentz reciprocity [41,42]. The correspondence between DDS and NHSE states that $\text{Im}[\omega^P(k_+)] = \text{Im}[\omega^P(k_-)]$ implies the absence of NHSE. By Eqs. (8) and (9), we have the criterion for the absence of NHSE

$$\Delta = -n''_{\text{eff}}\xi'_{\text{eff}} + n'_{\text{eff}}\xi''_{\text{eff}} = 0. \quad (10)$$

Equivalently, the occurrence of NHSE requires the non-zero $\Delta$. Together with Eqs. (5-7), Eq. (10) already determines the emergence of NHSE without invoking BCs. However, as a phenomenon unique in the OBC, the NHSE shall depend on the BCs, and thus, we will unveil next that the non-Bloch theory for the continuum system [33] gives the same criterion as Eq. (10).

**Section III. Non-Bloch theory of the effective medium**

We now focus on the localization behavior of eigenmodes under the OBC within the EMT description, where the eigenproblems for finite systems under different BCs can be solved analytically. Specifically, assume the effective medium defined by Eq. (4) is placed in a cavity with its boundaries being perfect electric conductors (PECs) applied at $x = 0$ and $x = L$, the eigenfrequency and eigenmodes are

$$\frac{\omega_m^{ee}}{c} = \frac{m\pi}{\sqrt{\varepsilon_{\text{eff},z}\mu_{\text{eff},y}}L} = \frac{m\pi}{L\left[\left(n'_{\text{eff}}\right)^2 + \left(n''_{\text{eff}}\right)^2\right]}\left(n'_{\text{eff}} - in''_{\text{eff}}\right), \tag{11}$$

$$E_z^{ee}(x) = 2iAe^{i\frac{\omega_m^{ee}}{c}\xi_{\text{eff}}x}\sin\left(\frac{\omega_m^{ee,\prime}}{c}n'_{\text{eff}}x - \frac{\omega_m^{ee,\prime\prime}}{c}n''_{\text{eff}}x\right), \tag{12}$$

$$H_y^{ee}(x) = -2\frac{A}{Z_{\text{eff}}}e^{i\frac{\omega_m^{ee}}{c}\xi_{\text{eff}}x}\cos\left(\frac{\omega_m^{ee,\prime}}{c}n'_{\text{eff}}x - \frac{\omega_m^{ee,\prime\prime}}{c}n''_{\text{eff}}x\right), \tag{13}$$

where $A$ is a normalization constant and $m = 1, 2, \cdots$. The superscript $ee$ signifies that both left and right boundaries are PECs. The above results show that the electromagnetic coupling term $\xi_{\text{eff}}$ here do not affect the OBC energy spectrum, which will be further clarified in Sec. IV. In the thermodynamic limit ($L \to \infty$), whether the eigenmodes are localized or not is determined by

$$\text{Im}\,\frac{\omega_m^{ee}}{c}\xi_{\text{eff}} = k_c(-n''_{\text{eff}}\xi'_{\text{eff}} + n'_{\text{eff}}\xi''_{\text{eff}}) = k_c\Delta, \tag{14}$$

where $k_c = \frac{m\pi}{L\left[\left(n'_{\text{eff}}\right)^2+\left(n''_{\text{eff}}\right)^2\right]} \in \mathbb{R}_+$ becomes continuous in the large-$L$ limit. The absence of NHSE requires $\text{Im}\,\frac{\omega_m^{ee}}{c}\xi_{\text{eff}} = 0$, leading to $\Delta = 0$, which is the same as Eq. (10). In Appendix C, we analytically provide the eigenfrequency and eigenmodes under different BCs. It is evident that, in the thermodynamic limit, the OBC spectra and localization behavior remain the same for different BCs. This is a characteristic of non-Hermitian continuum systems, where their bulk properties are entirely determined by the number of left-hand-side and right-hand-side BCs, with the specifics of these conditions only affecting the details of eigenmodes [33].

The above analysis provides us with the eigenfrequency and concrete form of the eigenmodes under the OBC in the thermodynamic limit. Since the essence of the non-Bloch band theory is that it is capable of predicting the OBC spectra in the thermodynamic limit from the unit cell by using the GBZ, we then need the non-Bloch theory for the continuum effective medium. Technically speaking, the GBZ is of the same dimension as the physical one, and thus, on top of vanishing the characteristic polynomial, another constraint is required to derive from the BCs in 1D. The number of BCs applied at each end is one, and the continuum GBZ condition is then $\mathrm{Im}\,k_+(\omega) = \mathrm{Im}\,k_-(\omega)$ [33,66]. By expressing $\mathrm{Im}\,k_\pm$ explicitly from Eq. (8) as

$$\mathrm{Im}\,k_\pm(\omega) = \left(\frac{\omega''}{c}\xi'_{\mathrm{eff}} + \frac{\omega'}{c}\xi''_{\mathrm{eff}} \pm \frac{\omega''}{c}n'_{\mathrm{eff}} \pm \frac{\omega'}{c}n''_{\mathrm{eff}}\right), \tag{15}$$

the condition $\mathrm{Im}\,k_+(\omega) = \mathrm{Im}\,k_-(\omega) = \tau(\omega)$ gives the following

$$\frac{\omega''}{c}n'_{\mathrm{eff}} + \frac{\omega'}{c}n''_{\mathrm{eff}} = 0, \tag{16}$$

$$\tau(\omega) = \frac{\omega''}{c}\xi'_{\mathrm{eff}} + \frac{\omega'}{c}\xi''_{\mathrm{eff}} = k_c(-n''_{\mathrm{eff}}\xi'_{\mathrm{eff}} + n'_{\mathrm{eff}}\xi''_{\mathrm{eff}}) = k_c\Delta, \tag{17}$$

where $k_c$ still stands for a continuous variable as Eq. (14). The quantity $e^{-\tau}$ here plays the same role as $|\beta|$ in the GBZ of lattice models. Besides yielding to the occurrence criteria of NHSE in Eq. (10), Eq. (17) also tells that the sign of $\tau(\omega)$ demarcates the localization characteristics of NHSE. Hence, the dispersion relation [Eq. (8)] shall be solved together with the continuum GBZ condition to identify $\tau(\omega)$ and the ensuing OBC spectra. The solid lines in Fig. 1(g) show the complex-$k$ BS by solving Eq. (8) for the purely real frequencies, and we see $\mathrm{Im}\,k_+(\omega) = \mathrm{Im}\,k_-(\omega) > 0$, voluntarily satisfying the continuum GBZ condition. This elucidates that the OBC spectra of such EMT lie in the real frequency axis, and the eigenmodes are the left-going skin modes.

Although the complex-$k$ BS in Fig. 1(g) is somewhat unique to fulfill the continuum GBZ condition, such perception is surely a general recipe to determine the OBC behavior of a continuum system, which supplements the DDS perception. As shown in Figs. 1(f) and 1(h), we consider a wave at a given frequency incident on the interface, and the reflected wave must also be at this frequency. The DDS argument states that the reflected wave must possess the same lifetime as the incident wave to make forming standing waves possible [Fig. 1(f)]. In complement, the continuum GBZ condition ensures that a reflected wave with the same localization as the incident wave exists. This allows their superposition to satisfy the BCs, thus forming the standing waves [Fig. 1(h)]. Hence, the continuum GBZ condition quantitatively decides the OBC behaviors instead of qualitatively predicting the occurrence and tendency of skin modes.

Since all the above considerations converge to the same criterion for the emergence of NHSE, we depict in Fig. 2 the distribution of $\Delta$ in different parameter planes. By considering $|\alpha_{\varepsilon,\mu}| \ll 1$ in reality, resulting in $n''_{\text{eff}} \ll n'_{\text{eff}}$ and $n'_{\text{eff}}\xi''_{\text{eff}} \gg n''_{\text{eff}}\xi'_{\text{eff}}$, we set $\phi_\varepsilon = 0$ and $\phi_\mu = 0$ throughout this work to maximize $\xi_{\text{eff}}$. Consequently, we investigate $\Delta$ in two parameter planes: $(\arg(\alpha_\varepsilon), \arg(\alpha_\mu))$ and $(g, \Omega)$. Since the dominant contribution is from $n'_{\text{eff}}\xi''_{\text{eff}}$, the maximum value of $\Delta$ in the $\arg(\alpha_\varepsilon) - \arg(\alpha_\mu)$ plane occurs at $\arg(\alpha_\varepsilon) + \arg(\alpha_\mu) = \pi/2 + n\pi$, consistent with Fig. 2(a). This relation, which the modulation phase of $\alpha_\varepsilon$ and $\alpha_\mu$ should satisfy to maximize $\Delta$, indicates that synchronously modulating imaginary parts of $\varepsilon$ and $\mu$ does not imply the NHSE. In contrast, modulating real parts of $\varepsilon$ ($\mu$) and imaginary parts of $\mu$ ($\varepsilon$) simultaneously will optimize the localization of skin modes, which is just the case shown in Fig. 1 [highlighted by the red markers in Fig. 2(a)]. Besides the magnitude of $\Delta$, Fig. 2(a) also shows its sign change by varying $\arg(\alpha_\varepsilon)$ and $\arg(\alpha_\mu)$, which is seemingly straightforward from Eq. (7). Whereas, the sign of $\Delta$ also flips when the modulation transits from the subluminal region ($\Omega < vg$) to the superluminal one ($\Omega > vg$) with all other parameters fixed. For the modulation at the red markers in Fig. 2(a), we show in Fig. 2(b) the distribution of $\Delta$ in the $g - \Omega$ plane. The sign change is clearly seen when the modulation goes across the speed of light ($\Omega = vg$), and the results near $\Omega = vg$ become vague because the EMT is invalid therein [67,68]. It is worth pointing out that the NHSE is absent when the modulation becomes purely spatial ($\Omega = 0$) or temporal ($g = 0$) since therein exists symmetry to make the spectra purely real or complex pairs, leading to the absence of point gaps [7-12,29,30].

Till now, we have expounded the fundamental Floquet bands in the long wavelength limit of the STM with their critical features under both PBC and OBC able to capture by the EMT faithfully. The power and conciseness of EMT afford a handy way toward scrutinizing the non-Hermitian system containing the STM as one component, which will be demonstrated next.

### Section IV. Generalized Brillouin zone of spatiotemporal photonic crystals

In order to corroborate the EMT recipe for the STM that can successfully embed in the composite non-Hermitian system, we now consider a spatial PhC composed of homogeneous materials and STMs, as illustrated in Fig. 3(a). The boundaries are still formed spatially, and thus, we dub the system in Fig. 3(a) the STM-PhC. Let us begin with solving the PBC spectra of this STM-PhC. Due to the periodicity in space-time, the modes still obey the form of Bloch-Floquet states, and the PWE method shown in Appendix A is then available to determine the PBC spectra, as displayed by the open stars in Fig. 3(b). Only the lower two bands are shown herein because $\mathrm{Re}(\omega\Lambda/2\pi c) \leq 0.6$ (equivalently, $\mathrm{Re}(\omega/\Omega) \leq 0.15$) is within the validity of EMT. This permits replacing the STM with an effective bianisotropic medium as component $b$, and now the PhC is constituted by two homogenous media, referred to as an effective photonic crystal (E-PhC). The corresponding E-PhC results are shown in Fig. 3(b) by the solid lines, and good agreement is seen compared with the PWE method. The color of the stars and lines denotes $\mathrm{Im}(\omega^P)$, and the correspondence from the DDS perspective indicates the NHSE for both bands propagating towards the left-hand side ($-x$ direction).

To validate the NHSE, we investigate a finite-sized STM-PhC by applying PEC BCs at both ends. The PEC here implies $E_z = 0$ and thus is equivalent to the Dirichlet BCs used in condensed matter. The OBC spectra generally relate to the scattering matrix, so we first formulate the TMM for the STM-PhC [64,69], and the scattering matrix is then obtained recursively from the transfer matrix (see Appendix D for details) [70]. Since the TMM is numerically unstable when the wavenumber becomes complex, or the system size increases, we adopt TMM for the unit-cell level calculations but use the scattering matrix method (SMM) for finite-sized calculations. Before obtaining the scattering matrix, we first apply the PBCs in the TMM to acquire the PBC band structure [circles in Fig. 3(b)], which agrees well with both PWE and E-PhC, validating the transfer matrix. By imposing the PEC BCs in the scattering matrix, we obtain the mode condition function $g(\omega)$ (see Appendix D for explicit expressions), which is plotted by the color contours in Fig. 3(c). The zeros of $g(\omega)$, which correspond to the OBC spectra, are highlighted by the red diamond markers, and two bands and one in-gap state are seen. The $\mathrm{Re}(\omega^O)$ range of both bands (the superscript $O$ stands for the OBC case) is almost the same as that in Fig. 3(b), but the $\mathrm{Im}(\omega^O)$ is one order of magnitude smaller than $\mathrm{Im}(\omega^P)$ in Fig. 3(b). Such disparateness hints at the NHSE, as demonstrated by the electric field distribution (the red dashed line) in Fig. 3(d) for the bulk state marked with the black pentagram in Fig. 3(c). The skin modes apparently localize at the left-hand side boundary, confirming the previous statements. For comparison, the electric field distribution for the in-gap state is depicted in Fig. 3(e) by the red dashed line, which is localized at the right-hand side boundary. The occurrence of NHSE and the in-gap state begs for the non-Bloch band theory because the GBZ informs the skin mode localization behavior and identifies the parametric loop on which the integral of the topological invariant performs.

As stated previously, the vanishing of the characteristic polynomial $f(\beta,\omega)=0$, where $\beta=e^{iq}$, is inadequate to determine the GBZ, whatever $f(\beta,\omega)$ is from the tight-binding model (TBM) or the TMM in the STM-PhC. Specifically in the 1D lattice model, another constraint derived from Dirichlet BCs is known as $|\beta_M(\omega)|=|\beta_{M+1}(\omega)|$. This states that for the $\omega$ in the OBC spectra, the norm of two middle roots of $\beta$ shall be equal when the $2M$ roots of $\beta$ are sorted by their moduli incrementally. Although the BCs in Fig. 3(a) are essentially Dirichlet BCs, we prove from the transfer matrix that the criterion for discriminating the OBC spectra in the STM-PhC is still

$$\left|\beta_{2l_c+1}(\omega)\right|=\left|\beta_{2l_c+2}(\omega)\right|. \tag{18}$$

Here, $\beta_i$ is the *i*-th eigenvalue of the transfer matrix. These eigenvalues are sorted in ascending order of their magnitudes as $|\beta_1|\le|\beta_2|\le\cdots\le|\beta_{(4l_c+1)}|\le\left|\beta_{(4l_c+2)}\right|$, where $l_c$ is the cutoff defined in the PWE (see Appendices A and E for details). Figures 4(a) and 4(b) show the GBZ calculated by Eq. (18) for the lower two bands (blue circles). The radii of both GBZs are smaller than one, indicating the skin modes localized at the left boundary, which is qualitatively the same as the continuum medium. We further obtain the corresponding $\omega^O$ on the GBZs, which is depicted in Fig. 3(c) by the blue lines. The consistency between the SMM and GBZ seen in Fig. 3(c) validates Eq. (18).

To validate the EMT description in the OBC, we recall the E-PhC setup. The OBC spectra and field distributions calculated by the SMM are represented by green dots and lines in Fig. 3(c-e), and the acquired GBZs by Eq. (18) with $l_c=0$ are shown by the green dashed lines in Fig. 4. Again, excellent agreement between the E-PhC and STM-PhC further manifests that the EMT is a faithful description and connotes the possibility of EMT-based GBZ being accessible analytically. The electromagnetic fields for $E_z$ polarization of the E-PhC satisfy

$$\hat{\boldsymbol{L}}\boldsymbol{\psi}(x)=\frac{\omega}{c}\boldsymbol{K}_{\mathrm{E}}(x)\boldsymbol{\psi}(x), \tag{19}$$

$$\hat{\boldsymbol{L}}=\begin{pmatrix}0 & i\dfrac{\mathrm{d}}{\mathrm{d}x}\\ i\dfrac{\mathrm{d}}{\mathrm{d}x} & 0\end{pmatrix},\boldsymbol{K}_{\mathrm{E}}(x)=\begin{pmatrix}\varepsilon_{\mathrm{E},z}(x) & -\xi_{\mathrm{E}}(x)\\ -\xi_{\mathrm{E}}(x) & \mu_{\mathrm{E},y}(x)\end{pmatrix},\boldsymbol{\psi}(x)=\begin{pmatrix}\sqrt{\varepsilon_0}E_z(x)\\ \sqrt{\mu_0}H_y(x)\end{pmatrix}, \tag{20}$$

where the subscript *E* in $\boldsymbol{K}_{\mathrm{E}}$ signifies that it is the material parameter matrix describing the E-PhC. The matrix components of $\boldsymbol{K}_{\mathrm{E}}$ are piecewise continuous functions. By considering the following transformation [59,60]

$$\boldsymbol{\psi}(x)=r(x)\boldsymbol{D}(x)\boldsymbol{Q}(x), \tag{21}$$

where $r(x)=\exp\left[i\int_0^x \frac{\omega}{c}\xi_{\mathrm{E}}(v)\mathrm{d}v\right]$, $\boldsymbol{D}(x)=\mathrm{Diag}\left[\left(\varepsilon_{\mathrm{E},z}(x)\right)^{-1/2},\left(\mu_{\mathrm{E,y}}(x)\right)^{-1/2}\right]$, and $\boldsymbol{Q}(x)=[Q_{\mathcal{E}}(x)\quad Q_{\mathcal{H}}(x)]^T$ is an auxiliary field quantity defined by Eq. (21), we reformulate Eq. (19) as

$$r(x)\hat{\boldsymbol{L}}\boldsymbol{D}(x)\boldsymbol{Q}(x)=\left[\frac{\omega}{c}\boldsymbol{K}_{\mathrm{E}}(x)r(x)-\left(\hat{\boldsymbol{L}}r(x)\right)\right]\boldsymbol{D}(x)\boldsymbol{Q}(x)=\frac{\omega}{c}r(x)\boldsymbol{D}(x)^{-1}\boldsymbol{Q}(x). \quad (22)$$

Subsequently, by introducing the operator $\hat{\boldsymbol{L}}_S=\boldsymbol{D}(x)\hat{\boldsymbol{L}}\boldsymbol{D}(x)$, we reorganize Eq. (22) into a more concise form as

$$\hat{\boldsymbol{L}}_S\boldsymbol{Q}(x)=\frac{\omega}{c}\boldsymbol{Q}(x). \quad (23)$$

The absence of $\xi_{\mathrm{E}}$ in the operator $\hat{\boldsymbol{L}}_S$ indicates that the E-PhC spectra are disconnected from $\xi_{\mathrm{E}}$, as exhibited in Eq. (11) and Appendix C, and either $Q_{\mathcal{E}}(x)$ or $Q_{\mathcal{H}}(x)$ can then be used to determine the E-PhC spectra. The PEC ($E_z=0$) and perfect magnetic conductor (PMC, $H_y=0$) BCs become $Q_{\mathcal{E}}=0$ and $Q_{\mathcal{H}}=0$ at the boundary, which are essentially Dirichlet or Neumann BCs, depending on the $Q_{\mathcal{E}}$ or $Q_{\mathcal{H}}$ being employed. All these conclude that the OBC spectra of the E-PhC [Eq. (19)] can be investigated from the operator $\hat{\boldsymbol{L}}_S$ [Eq. (23)].

Concerning Fig. 3, it is crucial that $\varepsilon_{\mathrm{E},z}(x)$ and $\mu_{\mathrm{E,y}}(x)$ are both real, thus ensuring the Hermiticity of the operator $\hat{\boldsymbol{L}}_S$ under the inner product $\langle\boldsymbol{Q_1}|\boldsymbol{Q_2}\rangle=\int_{x_1}^{x_2}(\boldsymbol{Q_1})^{\dagger}\boldsymbol{Q_2}dx$ [71], where the integration interval spans a unit cell (entire system) under the PBC (OBC). This explains why the OBC spectra of the E-PhC lie entirely on the real axis. What is more profound is that the GBZs of the E-PhC [Eq. (19)] can then be determined by Eqs. (21) and (23) analytically as

$$\beta_{n_i}(q)=\exp\left[i\int_{\mathrm{uc}}\frac{\omega_{n_i}^S(q)}{c}\xi_{\mathrm{E}}(v)\mathrm{d}v+iq\Lambda\right], \quad (24)$$

where $\omega_{n_i}^S(q)$ stands for the PBC spectra of $\hat{\boldsymbol{L}}_S$ with its subscript $n_i$ (superscript $S$) denoting the band index (the $\hat{\boldsymbol{L}}_S$ case). Considering $\xi_{\mathrm{E}}(x)$ is a piecewise function, we can then see that $\ln\left|\beta_{n_i}(q)\right|=-c^{-1}\omega_{n_i}^S(q)\xi_{\mathrm{E}}''f_b\Lambda$, where $f_b=d_b/\Lambda$ is the filling ratio of the STM. When $f_b=0$ ($f_b=1$), $\ln\left|\beta_{n_i}(q)\right|=0$ ($\ln\left|\beta_{n_i}(q)\right|=-\tau\Lambda$) recovers the BZ [the continuum GBZ in Eq. (17)]. This indicates that the GBZ radius of the E-PhC is determined by the filling ratio and dispersion relations together as shown by the solid red lines in Fig. 4. The excellent agreement herein reveals that Eq. (24) offers a straightforward approach to accessing the GBZ of the STM-PhC, and Eq. (21) plays exactly the same role of similarity transformation used in some intriguing TBMs [26]. Compared with Refs. [66,72], which also have utilized TMM to explore the non-Bloch properties of electromagnetic coupling materials, our method, say Eq. (21), not only provides analytical formulas for the GBZ but also eliminates the requirement of left eigenvectors when calculating topological invariants due to the Hermiticity of $\hat{\boldsymbol{L}}_S$. With such a powerful tool, we are now ready to investigate the topological properties of the STM-PhC.

**Section V. Localization and delocalization of topological edge states**

To examine the edge states of the STM-PhC system, we construct a domain wall formed by two STM-PhCs, as depicted in Fig. 5(a). The unit cells of both STM-PhCs are chosen to be symmetric, which will simplify the following analysis. To better leverage the analytical theory developed in the previous sections, we aim to reveal the topological properties of STM-PhCs from the E-PhC approach and verify the OBC spectra using the method in Appendix D. We choose the STM-PhC2 to the one already investigated in Fig. 3, and Fig. 5(b) shows its non-Bloch BS, which contrasts the Bloch BS, say Fig. 3(b). With the vertical axis still showing $\mathrm{Re}(\omega\Lambda/2\pi c)$, the horizontal axis uses the argument of $\beta$ instead of $q$, while the line color reflects $|\beta|$. Such non-Bloch BS with the GBZs and the OBC spectra information contained shall be utilized when investigating the topological properties of a finite-sized system. We focus on the second non-Bloch band gap, and the non-Bloch Zak phases of the two bands below the gap are indicated therein. The non-Bloch Zak phase is calculated on the GBZ by using the following biorthogonal Berry connection

$$\theta_{n_i}^{\mathrm{Zak}} = \oint_{\mathrm{GBZ}_{\mathrm{n_i}}} \mathrm{d}\varphi_\beta \left[ i \left\langle u_{n_i,\varphi_\beta}^{L} \middle| \partial_{\varphi_\beta} u_{n_i,\varphi_\beta}^{R} \right\rangle \right], \tag{25}$$

where $\varphi_\beta = \arg(\beta)$, $n_i$ is the band index, and $u^R$ ($u^L$) represents the periodic wave function of the non-Bloch right (left) wave function (see Appendix F for the details regarding $u^R$ and $u^L$). The transformation of Eq. (21) implies that the conclusions drawn from the Hermitian PhC defined by Eq. (23) can be applied to the STM-PhCs, which guarantees the quantization of the non-Bloch Zak phase shown in Fig. 5(b) [73-75].

To make TESs occur in the second gap, we introduce the STM-PhC1 by varying $\varepsilon_{\mathrm{E,z}}^{a1}$ and adjusting $\varepsilon_{\mathrm{E,z}}^{b1}$ accordingly to pin its non-Bloch band gap center at $\frac{2\pi c}{(n_{\mathrm{E}}^{a2} d_{a2} + n_{\mathrm{E}}^{b2} d_{b2})}$, where $\{a1, a2, b1, b2\}$ denotes the regions claimed in Fig. 5(a) and $n_E = \sqrt{\varepsilon_{\mathrm{E},z}\mu_{\mathrm{E},y}}$. All other parameters of the STM-PhC1 are fixed and claimed in the caption of Fig. 5. Figure 5(c) depicts the OBC spectra of the STM-PhC1 by the grey shaded area, and the non-Bloch Zak phase of the second non-Bloch band is depicted by the solid green line. It is evident that the non-Bloch Zak phase successfully coincides with the non-Bloch band gap closure at $\varepsilon_{\mathrm{E,z}}^{a1} \approx 1.51$. Its PBC spectra depicted by the light yellow shaded area are shown in Fig. 5(d) for comparison, where the second Bloch band gap is closed within $\varepsilon_{\mathrm{E,z}}^{a1} = 1.39{\sim}1.65$. Such inconsistency between the PBC and OBC results underscores the importance of the GBZ in non-Hermitian topology. To further illustrate, we then calculate the non-Bloch BS and Zak phases of the E-PhC1 at $\varepsilon_{\mathrm{E,z}}^{a1} = 1.2$ and $\varepsilon_{\mathrm{E,z}}^{a1} = 1.9$, as shown in the Figs. 5(e) and 5(f). The non-Bloch Zak phase of the first band is $\pi$ and will always be $\pi$ within the chosen $\varepsilon_{\mathrm{E,z}}^{a1}$ range. Therefore, the TESs do not appear in the second non-Bloch band gap when $\varepsilon_{\mathrm{E,z}}^{a1} = 1.2$ but will emerge when $\varepsilon_{\mathrm{E,z}}^{a1} = 1.9$, as demonstrated by the OBC spectra in Figs. 5(g) and 5(h). The TESs are indeed observed when $\varepsilon_{\mathrm{E,z}}^{a1} = 1.9$, confirming that the developed non-Bloch theory can reflect the topological properties of STM-PhCs.

Another intriguing example in the non-Bloch band theory is that the distinct localization behavior of the TESs and skin modes can lead to the delocalization of the TES [66,76-78]. Since the skin modes previously all tend to the left ($|\beta| < 1$) and the TES in Fig. 3 localizes at the right-hand side, we choose the configuration in Fig. 3(a). The PBC and OBC spectra in the Hermitian scenario are first shown in Fig. 6(a), and the TES (one bulk state) is highlighted by a blue circle (red circle) with their mode profiles shown in Fig. 6(c). To delocalize the TES, we shall turn on the spatiotemporal modulation to reach an appropriate value of $\xi_{\mathrm{E}}$, which compensates for the decay of TESs when performing the transformation in the GBZ. Figure 6(b) exhibits the PBC (yellow lines) and OBC (circles) spectra when $\alpha_{\varepsilon} = 0.1915i$ and $\alpha_{\mu} = 0.1915$. The OBC spectra shrink from the PBC ones, hinting at the NHSE, as shown in the top panel of Fig. 6(d). However, the TES (the blue circle) now resides on the PBC spectra and thus becomes delocalized, as illustrated by its mode profile in the bottom panel of Fig. 6(d). Here in Figs. 5 and 6, the employment of the GBZ established in Sec. IV validates the non-Bloch BS for the STM-PhC, thus providing a recipe for analyzing the fundamental Floquet bands when they experience non-Hermiticity.

## Section VI. Discussions and Conclusions

In summary, we have established the non-Bloch band theory for the $(1 + 1)$-dimensional PhCs with the aid of the effective medium description for the STM. Due to the fact that the STM in the long-wavelength limit is able to be modeled by the effective medium, and also thanks to the DDS viewpoint and continuum GBZ condition, we have firstly analytically made clear the occurrence condition and mode profile of NHSE in the STMs. Based on such an EMT recipe, the non-Bloch band theory for the STM-PhCs is then successfully formulated and verified by the TMM with multiple Floquet sidebands included. As a consequence, the GBZ obtained from the non-Bloch band theory has demonstrated that it can predict the topological transition of bulk bands, restore the bulk-boundary correspondence, and realize the delocalization of TESs. The validity of our non-Bolch band theory only relies on the accuracy of effective parameters for the STM, so it can handle multiple bands and shall find its role in the non-Abelian Floquet system [79,80]. Concerning the generalization to $d \geq 2$ setups, if the EMT still works, and so does our prescription, which paves an alternative way to higher-dimensional spatiotemporal crystals. All powerful theoretical methods and fancy wave phenomena previously investigated at the EMT level can then be utilized to digest the non-Hermitian physics in higher-dimensional Floquet systems [80]. The holistic view of the $(d+1)$-dimensional problem requires further investigation because the space-time symmetry and spatiotemporal boundaries shall be considered together [81,82], but the numerical method established here provides one scheme to cope with both the spectra and wavefunctions.

**Acknowledgment**

We thank Dr. Jing Lin and Dr. Mengying Hu for the helpful discussions. This work is supported by the National Natural Science Foundation of China (12174072, 2021hwyq05), the National Key R&D Program of China (2022YFA1404500, 2022YFA1404701), and the Natural Science Foundation of Shanghai (No. 21ZR1403700).

## Appendix A: Plane wave expansion method

As a complete basis of wave equations, the expansion by plane waves is always one choice to calculate the BSs. To handle both the STM (Fig. 1) and the STM-PhC (Fig. 3), we re-express the permittivity in Eq. (2) as

$$\varepsilon(x,t) = \varepsilon_r(x)[1 + 2\alpha_\varepsilon(x)\cos(gx - \Omega t + \phi_\varepsilon)], \tag{A1}$$

$$\varepsilon_r(x) = \begin{cases} \varepsilon_a, & j\Lambda < x \le j\Lambda + d_a \\ \varepsilon_b, & j\Lambda + d_a < x \le (j+1)\Lambda \end{cases}, \tag{A2}$$

$$\alpha_\varepsilon(x) = \begin{cases} 0, & j\Lambda < x \le j\Lambda + d_a \\ \alpha_\varepsilon, & j\Lambda + d_a < x \le (j+1)\Lambda \end{cases}. \tag{A3}$$

The symbols $j$, $a$, and $b$ respectively denote the $j$-th unit cell, the component $a$, and the component $b$ depicted in Fig. 3, wherein $d_a$ and $\Lambda$ are also defined. Due to the spatial periodicity of $\varepsilon_r(x)$ and $\alpha_\varepsilon(x)$, $\varepsilon(x,t)$ can then formally be rewritten as

$$\varepsilon(x,t) = \sum_p \tilde{\varepsilon}_{r,p}\, e^{ipGx} + \sum_{p,o} \tilde{\varepsilon}_{r,p}\tilde{\alpha}^{+}_{\varepsilon,o} e^{i[(p+o+N)Gx-\Omega t]}$$
$$+ \sum_{p,o} \tilde{\varepsilon}_{r,p}\tilde{\alpha}^{-}_{\varepsilon,o} e^{-i[(-p-o+N)Gx-\Omega t]}, \tag{A4}$$

where the subscript $p$ $(o)$ of $\tilde{\varepsilon}_{\mathrm{r},p}$ $(\tilde{\alpha}^{\pm}_{\varepsilon,o} = \tilde{\alpha}_{\varepsilon,o} e^{\pm i\phi_\varepsilon})$ is the Fourier order running from $-\infty$ to $+\infty$, the tilded symbols denote the Fourier transformed quantity, and $G = g/N = 2\pi/\Lambda$. The form of Eqs. (A1–A4) also holds for the permeability by substituting $\varepsilon$ with $\mu$. Here, we focus on the $E_z$ polarization, and the electromagnetic fields shall satisfy

$$\frac{\partial\sqrt{\varepsilon_0}E_z(x,y,t)}{\partial y} = -\frac{1}{c}\frac{\partial}{\partial t}\mu(x,t)\sqrt{\mu_0}H_x(x,y,t), \tag{A5}$$

$$\frac{\partial\sqrt{\varepsilon_0}E_z(x,y,t)}{\partial x} = \frac{1}{c}\frac{\partial}{\partial t}\mu(x,t)\sqrt{\mu_0}H_y(x,y,t), \tag{A6}$$

$$\frac{\partial\sqrt{\mu_0}H_y(x,y,t)}{\partial x} - \frac{\partial\sqrt{\mu_0}H_x(x,y,t)}{\partial y} = \frac{1}{c}\frac{\partial}{\partial t}\varepsilon(x,t)\sqrt{\varepsilon_0}E_z(x,y,t). \tag{A7}$$

Due to the spatiotemporal periodicity, the field solution should possess the following form

$$\Phi(x,y,t) = \sum_{l,n=-\infty}^{\infty} \tilde{\Phi}_{ln} e^{i[k+(lN+n)G]x+ik_y y-i(\omega+l\Omega)t}, \tag{A8}$$

where $\Phi$ represents any quantity within $\sqrt{\varepsilon_0}E_z$, $\sqrt{\mu_0}H_x$, and $\sqrt{\mu_0}H_y$, $k$ and $\omega$ are the wavenumber and Floquet frequency. Substituting Eqs. (A4) and (A8) into Eqs. (A5–A7), we obtain the following eigenvalue equation

$$k\begin{pmatrix}\tilde{\boldsymbol{E}}\\ \tilde{\boldsymbol{H}}\end{pmatrix} = \begin{pmatrix}\mathbf{M}^{EE} & \mathbf{M}^{EH}\\ \mathbf{M}^{HE} & \mathbf{M}^{HH}\end{pmatrix}\begin{pmatrix}\tilde{\boldsymbol{E}}\\ \tilde{\boldsymbol{H}}\end{pmatrix}, \tag{A9}$$

in which $\widetilde{\boldsymbol{E}}$ ($\widetilde{\boldsymbol{H}}$) is the column vector composed of the Fourier coefficients $\sqrt{\varepsilon_0}\tilde{E}_{z,ln}$ ($\sqrt{\mu_0}\tilde{H}_{y,ln}$) and the matrix elements of $\mathbf{M}$ are

$$\mathrm{M}^{EE}_{l'n',ln} = \mathrm{M}^{HH}_{l'n',ln} = -(lN+n)G\delta_{l',l}\delta_{n',n}, \quad \text{(A10)}$$

$$\mathrm{M}^{EH}_{l'n',ln} = -\left\{\begin{array}{c} (k_\omega + lk_\Omega)\tilde{\mu}_{r,n'-n}\delta_{l',l} + \\ [k_\omega + (l+1)k_\Omega]\left[\sum_{o=-\infty}^{\infty}\tilde{\mu}_{r,n'-(n+o)}\tilde{\alpha}^{+}_{\mu,o}\right]\delta_{l',l+1} + \\ [k_\omega + (l-1)k_\Omega]\left[\sum_{o=-\infty}^{\infty}\tilde{\mu}_{r,n'-(n+o)}\tilde{\alpha}^{-}_{\mu,o}\right]\delta_{l',l-1} \end{array}\right\}, \quad \text{(A11)}$$

$$\mathrm{M}^{HE}_{l'n',ln} = -\left\{\begin{array}{c} (k_\omega + lk_\Omega)\tilde{\varepsilon}_{r,n'-n}\delta_{l',l} - k_y^2(M^{EH})^{-1}_{l'n',ln} \\ [k_\omega + (l+1)k_\Omega]\left[\sum_{o=-\infty}^{\infty}\tilde{\varepsilon}_{r,n'-(n+o)}\tilde{\alpha}^{+}_{\varepsilon,o}\right]\delta_{l',l+1} + \\ [k_\omega + (l-1)k_\Omega]\left[\sum_{o=-\infty}^{\infty}\tilde{\varepsilon}_{r,n'-(n+o)}\tilde{\alpha}^{-}_{\varepsilon,o}\right]\delta_{l',l-1} \end{array}\right\}. \quad \text{(A12)}$$

The $k_\omega$ ($k_\Omega$) in Eqs. (A11-A12) is defined as $\omega/c$ ($\Omega/c$). In principle, the matrix dimension and the series expansion shall take to infinity. However, each index must be given a cutoff numerically as $l_c$, $n_c$, and $o_c$. The dimension of $\mathbf{M}$ is $2(2n_c+1)(2l_c+1)$, and the summation in Eqs. (A11-A12) runs from $-o_c$ to $+o_c$. To acquire converged results, the values of $l_c$, $n_c$, and $o_c$ need to be meticulously chosen.

**Appendix B: Effective medium parameters of the spatiotemporal material**

Deriving the effective medium parameters of STM is firstly delivered by the conventional approach, and we further elucidate the spirit of EMT from the averaging fields in temporal domains. Since $\varepsilon_r(x)$ and $\alpha_\varepsilon(x)$ are both constants spatially for STM, only zeroth-order Fourier components are non-zero, i.e., $\tilde{\varepsilon}_{r,p} = \varepsilon_r\delta_{p,0}$ and $\tilde{\alpha}^{\pm}_{\varepsilon,o} = \alpha^{\pm}_{\varepsilon}\delta_{o,0}$. Substituting them into Eqs. (A10–A12), we can see that different $n$ components decouple, and thus, only the $n = 0$ set of equations needs to be considered to obtain the low-frequency dispersion relations. To the leading correction, we consider the influence of the nearest sidebands ($l = \pm 1$) on the zeroth-order energy band ($l = 0$), and Eq. (A9) becomes an eigenvalue equation for a 6x6 matrix, from which the analytical expressions of the zeroth-order band shall be then derived [44]. To explicate this process, we redefine the basis vectors as $\left(\sqrt{\varepsilon_r}\widetilde{\boldsymbol{E}} + \sqrt{\mu_r}\widetilde{\boldsymbol{H}}, \sqrt{\varepsilon_r}\widetilde{\boldsymbol{E}} - \sqrt{\mu_r}\widetilde{\boldsymbol{H}}\right)^T$, and Eq. (A9) is expressed as

$$k\begin{pmatrix}\sqrt{\varepsilon_r}\tilde{\boldsymbol{E}}+\sqrt{\mu_r}\tilde{\boldsymbol{H}}\\ \sqrt{\varepsilon_r}\tilde{\boldsymbol{E}}-\sqrt{\mu_r}\tilde{\boldsymbol{H}}\end{pmatrix}=\begin{pmatrix}\mathbf{M}^{++} & \mathbf{M}^{+-}\\ \mathbf{M}^{-+} & \mathbf{M}^{--}\end{pmatrix}\begin{pmatrix}\sqrt{\varepsilon_r}\tilde{\boldsymbol{E}}+\sqrt{\mu_r}\tilde{\boldsymbol{H}}\\ \sqrt{\varepsilon_r}\tilde{\boldsymbol{E}}-\sqrt{\mu_r}\tilde{\boldsymbol{H}}\end{pmatrix}. \tag{B1}$$

Considering the long-wavelength approximation ($k \ll g$ and $\omega \ll \Omega$) and small modulation amplitudes ($|\alpha_\epsilon| \ll 1$ and $|\alpha_\mu| \ll 1$), we obtain the quadratic equation in $k$

$$\beta^2\omega^2 = \kappa^2 k_y^2 + (k-\delta\omega)^2, \tag{B2}$$

where

$$\beta^2 = v^{-2}\left(1+\alpha_\varepsilon^+\alpha_\varepsilon^-\frac{2\Omega^2}{v^2g^2-\Omega^2}\right)\left(1+\alpha_\mu^+\alpha_\mu^-\frac{2\Omega^2}{v^2g^2-\Omega^2}\right), \tag{B3}$$

$$\kappa^2 = \left(1+\alpha_\mu^+\alpha_\mu^-\frac{2\Omega^2}{v^2g^2-\Omega^2}\right), \tag{B4}$$

$$\delta = \left(\alpha_\varepsilon^+\alpha_\mu^- + \alpha_\varepsilon^-\alpha_\mu^+\right)\frac{\mathrm{g}\Omega}{v^2g^2-\Omega^2} = 2\alpha_\varepsilon\alpha_\mu\cos\left(\phi_\varepsilon-\phi_\mu\right)\frac{\mathrm{g}\Omega}{v^2g^2-\Omega^2}. \tag{B5}$$

By comparing Eq. (B2) and the bianisotropic material described by Eq. (4), we can obtain the effective medium parameters as given in Eqs. (5–8). Although Eq. (B2) holds for non-zero $k_y$, we only consider the case $k_y = 0$ subsequently.

We then inspect the electromagnetic fields by comparing PWE and EMT to validate EMT further. The EMT is essentially an averaging field theory, and antecedent attempts mainly deal with spatial averaging, but the EMT above also performs time averaging. To unveil this, we express the electromagnetic fields in STM as a superposition of the eigenmodes given in Eq. (A8)

$$\begin{aligned} E_z(x,t) = &\sum_{\sigma_+} A^{k_{+,\sigma_+}}\tilde{E}_{l=0}^{k_{+,\sigma_+}}\, e^{ik_{+,\sigma_+}x-i\omega t} + \sum_{\sigma_-} B^{k_{-,\sigma_-}}\tilde{E}_{l=0}^{k_{-,\sigma_-}}\, e^{ik_{-,\sigma_-}x-i\omega t} \\ &+\sum_{\sigma_+} A^{k_{+,\sigma_+}}\sum_{l\neq 0}\tilde{E}_l^{k_{+,\sigma_+}}\, e^{i(k_{+,\sigma_+}+lg)x-i(\omega+l\Omega)t} \\ &+\sum_{\sigma_-} B^{k_{-,\sigma_-}}\sum_{l\neq 0}\tilde{E}_l^{k_{-,\sigma_-}}\, e^{i(k_{-,\sigma_-}+lg)x-i(\omega+l\Omega)t}. \end{aligned} \tag{B6}$$

There are a total of $2(2l_c+1)$ wavenumbers from Eq. (A9), and we divide them into two subsets: $\{k_{+,\sigma_+=-l_c,\cdots,0,\cdots,+l_c}\}$ and $\{k_{-,\sigma_-=-l_c,\cdots,0,\cdots,+l_c}\}$, based on their energy flow along the positive or negative $x$-direction. Each subset contains $(2l_c+1)$ wavenumbers denoted as $k_{+,\sigma_+/-,\sigma_-}$. The subscript $\sigma_\pm$ is the indices for each subset. $A^{k_{+,\sigma_+}}$ and $B^{k_{-,\sigma_-}}$ are the superposition coefficients determined by the boundary or initial conditions. Considering the long-wavelength case (ω ≪ Ω) and performing the time average over the period defined by Ω, i.e., $\langle E_z(x,t)\rangle = \frac{\Omega}{2\pi}\int_0^{2\pi/\Omega}\mathrm{d}t\, E_z(x,t)$, we obtain

$$\langle E_z(x,t)\rangle = \left[\sum_{\sigma_+} A^{k_{+,\sigma_+}} \tilde{E}_{l=0}^{k_{+,\sigma_+}}\, e^{ik_{+,\sigma_+}x} + \sum_{\sigma_-} B^{k_{-,\sigma_-}} \tilde{E}_{l=0}^{k_{-,\sigma_-}}\, e^{ik_{-,\sigma_-}x}\right] e^{-i\omega t}. \tag{B7}$$

Equation (B7) indicates that only the zeroth-order component ($l = 0$) is retained. Because the contribution from the sidebands to the slow-varying electromagnetic fields is relatively minor and we focus on the $k \to 0$ and $\omega \to 0$ scenario, Eq. (B7) can be further approximated as

$$\langle E_z(x,t)\rangle = \left[A^{k_{+,0}} \tilde{E}_{l=0}^{k_{+,0}} e^{ik_{+,0}x} + B^{k_{-,0}} \tilde{E}_{l=0}^{k_{-,0}} e^{ik_{-,0}x}\right] e^{-i\omega t}. \tag{B8}$$

From Eq. (B8), it is evident that the time-averaged field at low frequencies neglects higher-order components in the PWE (spatially and temporally fast-varying components), retaining only the lowest-order component (the slow-varying parts), which embodies the fundamental idea of EMT. The consistency between the electromagnetic fields of EMT and PWE, as shown in Fig. 3, validates the preceding discussion and contributes to a more comprehensive understanding of EMT established here.

## Appendix C: Spectra and eigenmodes under different boundary conditions

To supplement Sec. III, we provide the analytical OBC spectra and eigenmode profile by applying distinct BCs at $x = 0$ and $x = L$. With the PEC at $x = 0$ and PMC at $x = L$, the OBC spectra and eigenmodes are

$$\frac{\omega_m^{eh}}{c} = \frac{\left(m + \frac{1}{2}\right)\pi}{L\left[\left(n'_{\text{eff}}\right)^2 + \left(n''_{\text{eff}}\right)^2\right]}\left(n'_{\text{eff}} - i n''_{\text{eff}}\right), \tag{C1}$$

$$E_z^{eh} = 2iAe^{i\frac{\omega_m^{eh}}{c}\xi_{\text{eff}}x} \sin\left(\frac{\omega_m^{eh,\prime}}{c} n'_{\text{eff}}x - \frac{\omega_m^{eh,\prime\prime}}{c} n''_{\text{eff}}x\right), \tag{C2}$$

$$H_z^{eh} = -\frac{2A}{Z_{\text{eff}}} e^{i\frac{\omega_m^{eh}}{c}\xi_{\text{eff}}x} \cos\left(\frac{\omega_m^{eh,\prime}}{c} n'_{\text{eff}}x - \frac{\omega_m^{eh,\prime\prime}}{c} n''_{\text{eff}}x\right). \tag{C3}$$

The first (second) letter *e* (*h*) in the superscript *eh* indicates the left (right) boundary is the PEC (PMC). With the PMCs at both $x = 0$ and $x = L$, the OBC spectra and eigenmodes are

$$\frac{\omega_m^{hh}}{c} = \frac{m\pi}{L\left[\left(n'_{\text{eff}}\right)^2 + \left(n''_{\text{eff}}\right)^2\right]}\left(n'_{\text{eff}} - i n''_{\text{eff}}\right), \tag{C4}$$

$$E_z^{hh} = 2Ae^{i\frac{\omega_m^{hh}}{c}\xi_{\text{eff}}x} \cos\left(\frac{\omega_m^{hh,\prime}}{c} n'_{\text{eff}}x - \frac{\omega_m^{hh,\prime\prime}}{c} n''_{\text{eff}}x\right), \tag{C5}$$

$$H_z^{hh} = -\frac{2iA}{Z_{\text{eff}}} e^{i\frac{\omega_m^{hh}}{c}\xi_{\text{eff}}x} \sin\left(\frac{\omega_m^{hh,\prime}}{c} n'_{\text{eff}}x - \frac{\omega_m^{hh,\prime\prime}}{c} n''_{\text{eff}}x\right). \tag{C6}$$

With the PMC at $x = 0$ and PEC at $x = L$, the OBC spectra and eigenmodes are

$$\frac{\omega_m^{he}}{c} = \frac{\left(m + \frac{1}{2}\right)\pi}{L\left[\left(n'_{\text{eff}}\right)^2 + \left(n''_{\text{eff}}\right)^2\right]}\left(n'_{\text{eff}} - in''_{\text{eff}}\right), \tag{C7}$$

$$E_z^{he} = 2Ae^{i\frac{\omega_m^{he}}{c}\xi_{\text{eff}}x}\cos\left(\frac{\omega_m^{he,\prime}}{c}n'_{\text{eff}}x - \frac{\omega_m^{he,\prime\prime}}{c}n''_{\text{eff}}x\right), \tag{C8}$$

$$H_z^{he} = -\frac{2iA}{Z_{\text{eff}}}e^{i\frac{\omega_m^{he}}{c}\xi_{\text{eff}}x}\sin\left(\frac{\omega_m^{he,\prime}}{c}n'_{\text{eff}}x - \frac{\omega_m^{he,\prime\prime}}{c}n''_{\text{eff}}x\right). \tag{C9}$$

It is worth noting that when $L$ is finite, the OBC spectra are disparate under different BCs, but their difference $\frac{\pi/2}{L\left[\left(n'_{\text{eff}}\right)^2+\left(n''_{\text{eff}}\right)^2\right]}\left(n'_{\text{eff}} - in''_{\text{eff}}\right)$ tends to zero in the thermodynamic limit. In terms of the eigenmodes, the localization strength is entirely represented by the prefactor $e^{i\frac{\omega}{c}\xi x}$ of the electromagnetic fields.

### Appendix D: Transfer matrix method and scattering matrix

Establishing the transfer matrix requires knowing the eigenmodes of individual materials, and thus, we use the PWE method to obtain the eigenmodes of an STM. As stated in Appendix B, $\varepsilon_r$ and $\alpha_\varepsilon$ are both constants for the STM, and only the $n = 0$ equation set needs to be considered. The total fields within the vacuum and STM (denoted as $a$ and $b$ in the superscript) are then expanded as

$$E_z^{a/b}(x,t) = \sum_{\sigma_+} A_j^{k_{+,\sigma_+}^{a/b}} \sum_l \sqrt{\varepsilon_0}\tilde{E}_{z,l}^{k_{+,\sigma_+}^{a/b}} e^{i\left(k_{+,\sigma_+}^{a/b}+lg\right)\bar{x}-i(\omega+l\Omega)t} + \sum_{\sigma_-} B_j^{k_{-,\sigma_-}^{a/b}} \sum_l \tilde{E}_{z,l}^{k_{-,\sigma_-}^{a/b}} e^{i\left(k_{-,\sigma_-}^{a/b}+lg\right)\bar{x}-i(\omega+l\Omega)t}, \tag{D1}$$

$$H_y^{a/b}(x,t) = \sum_{\sigma_+} A_j^{k_{+,\sigma_+}^{a/b}} \sum_l \sqrt{\varepsilon_0}\tilde{H}_{y,l}^{k_{+,\sigma_+}^{a/b}} e^{i\left(k_{+,\sigma_+}^{a/b}+lg\right)\bar{x}-i(\omega+l\Omega)t} + \sum_{\sigma_-} B_j^{k_{-,\sigma_-}^{a/b}} \sum_l \tilde{H}_{y,l}^{k_{-,\sigma_-}^{a/b}} e^{i\left(k_{-,\sigma_-}^{a/b}+lg\right)\bar{x}-i(\omega+l\Omega)t}. \tag{D2}$$

Here, $A_j^{k_{+,\sigma_+}^{a/b}}$ ($B_j^{k_{-,\sigma_-}^{a/b}}$) is the expansion coefficient of the $k_{+,\sigma_+}^{a/b}$ ($k_{-,\sigma_-}^{a/b}$) eigenmode within the $a/b$ component of $j$-th unit cell. Besides, we have used $\bar{x} = x - (j-1)\Lambda$, where the $j$-th unit cell spans from $x = (j-1)\Lambda$ to $x = j\Lambda$. The continuity of electromagnetic fields imposes the following relation at the material interfaces

$$E_z^a[(j-1)\Lambda + d_a, t] = E_z^b[(j-1)\Lambda + d_a, t], \tag{D3}$$

$$H_y^a[(j-1)\Lambda + d_a, t] = H_y^b[(j-1)\Lambda + d_a, t], \tag{D4}$$

$$E_z^b[(j-1)\Lambda + d_a + d_b, t] = E_z^a[(j-1)\Lambda + d_a + d_b, t], \tag{D5}$$

$$H_y^b[(j-1)\Lambda + d_a + d_b, t] = H_y^a[(j-1)\Lambda + d_a + d_b, t]. \tag{D6}$$

Substituting Eqs. (D1-D2) into (D3-D6) leads to the explicit form of the transfer matrix

$$\begin{pmatrix} \boldsymbol{A}_{j+1}^a \\ \boldsymbol{B}_{j+1}^b \end{pmatrix} = \mathbf{T}(\omega) \begin{pmatrix} \boldsymbol{A}_j^a \\ \boldsymbol{B}_j^a \end{pmatrix}, \tag{D7}$$

where $\boldsymbol{A}_j^a$ ($\boldsymbol{B}_j^a$) is the column vector composed by $A_j^{k_{+,\sigma+}^a}$ ($B_j^{k_{-,\sigma-}^a}$) denoting the expansion coefficients at $x = (j-1)\Lambda^+$. The $\mathbf{T}(\omega)$ matrix contains the following five matrices as

$$\mathbf{T}(\omega) = \mathbf{P}(g, \Lambda)\boldsymbol{t}^{b\to a}\mathbf{P}(b, d_b)\boldsymbol{t}^{a\to b}\mathbf{P}(a, d_a), \tag{D8}$$

$$\boldsymbol{t}^{a\to b} = (\boldsymbol{t}^{b\to a})^{-1} = (\mathbf{V}^{\boldsymbol{b}})^{-\mathbf{1}}\mathbf{V}^{\boldsymbol{a}}, \tag{D9}$$

$$\mathbf{V}^{\boldsymbol{a}} = \left(\cdots, \begin{pmatrix} \widetilde{\boldsymbol{E}} \\ \widetilde{\boldsymbol{H}} \end{pmatrix}_{k_{+,\sigma+}^a}, \cdots, \begin{pmatrix} \widetilde{\boldsymbol{E}} \\ \widetilde{\boldsymbol{H}} \end{pmatrix}_{k_{-,\sigma-}^a}, \cdots\right), \tag{D10}$$

$$\mathbf{V}^{\boldsymbol{b}} = \left(\cdots, \begin{pmatrix} \widetilde{\boldsymbol{E}} \\ \widetilde{\boldsymbol{H}} \end{pmatrix}_{k_{+,\sigma+}^b}, \cdots, \begin{pmatrix} \widetilde{\boldsymbol{E}} \\ \widetilde{\boldsymbol{H}} \end{pmatrix}_{k_{-,\sigma-}^b}, \cdots\right), \tag{D11}$$

$$\mathbf{P}(a, d_a) = \begin{pmatrix} \mathrm{diag}(\ldots \quad e^{ik_{+,\sigma+}^a d_a} \quad \ldots) & \\ & \mathrm{diag}(\ldots \quad e^{ik_{-,\sigma-}^a d_a} \quad \ldots) \end{pmatrix}, \tag{D12}$$

$$\mathbf{P}(b, d_b) = \begin{pmatrix} \mathrm{diag}(\ldots \quad e^{ik_{+,\sigma+}^b d_b} \quad \ldots) & \\ & \mathrm{diag}(\ldots \quad e^{ik_{-,\sigma-}^b d_b} \quad \ldots) \end{pmatrix}, \tag{D13}$$

$$\mathbf{P}(g, \Lambda) = \begin{pmatrix} \mathrm{diag}(e^{-il_c g\Lambda} \quad \ldots \quad e^{il_c g\Lambda}) & \\ & \mathrm{diag}(e^{-il_c g\Lambda} \quad \ldots \quad e^{il_c g\Lambda}) \end{pmatrix}. \tag{D14}$$

$\mathbf{P}(a, d_a)$ and $\mathbf{P}(b, d_b)$ are the propagating matrix in components $a$ and $b$, respectively, and $\boldsymbol{t}^{a\to b}$ denotes the interface transfer matrix from component $a$ to $b$. Furthermore, $\mathbf{P}(g, \Lambda)$ is an additional phase matrix and returns to an identity matrix when $N$ is an integer, which is precisely our case. The symbols $k_{+,\sigma_+}^{a/b}$ and $(\widetilde{\boldsymbol{E}} \quad \widetilde{\boldsymbol{H}})_{k_{+,\sigma_+}^{a/b}}^T$, respectively, denote the eigenvalue and eigenvector of Eq. (A9), and the matrix $\mathbf{V}^{a/b}$ is composed of the eigenvector $(\widetilde{\boldsymbol{E}} \quad \widetilde{\boldsymbol{H}})_{k_{+,\sigma_+}^{a/b}}^T$.

The PBC BS is then ready to calculate by considering the following BCs as

$$E_z^a[j\Lambda, t] = \beta E_z^a[(j-1)\Lambda, t], \tag{D15}$$

$$H_y^a[j\Lambda, t] = \beta H_y^a[(j-1)\Lambda, t]. \tag{D16}$$

If we fix $\beta = e^{iq\Lambda}$, where $q$ is the Bloch wave vector, the above BC reverts to PBC. However, we still use $\beta$ for the subsequent generalization to the non-Bloch BS. By utilizing Eqs. (D1-D2), the above BCs (D15-D16) become

$$\begin{pmatrix} \boldsymbol{A}_{j+1}^{a} \\ \boldsymbol{B}_{j+1}^{a} \end{pmatrix} = \beta \begin{pmatrix} \boldsymbol{A}_{j}^{a} \\ \boldsymbol{B}_{j}^{a} \end{pmatrix}. \tag{D17}$$

Combining Eqs. (D7) and (D17) with fixed $\beta = e^{iq\Lambda}$, the PBC BS is determined by the following characteristic polynomial

$$f(\omega, q) = \det\left[\mathbf{T}(\omega) - e^{iq\Lambda}\right] = 0. \tag{D18}$$

By employing the same cutoff $l_c$, the dimensions of $\mathbf{t}^{a\to b}$, $\mathbf{P}(a, d_a)$, $\mathbf{T}(\omega)$, and similar quantities are $2(2l_c + 1)$. It is noteworthy that $\beta$ is the eigenvalue of $\mathbf{T}(\omega)$ and determines the propagation of the certain eigenfield after translating one unit cell. From this point, we can proceed to construct a non-Bloch band theory, and the detailed discussions are shown in Appendix E.

Since the propagating matrix involves $\infty$ and 0 when $k_{\pm,\sigma_\pm}^{a/b}$ becomes complex, the $\mathbf{T}(\omega)$ matrix is unstable numerically, and we adopt the recursive SMM to solve the problem under OBC [70]. The expansion coefficients of the first unit cell $(\boldsymbol{A}_1^a \quad \boldsymbol{B}_1^a)^T$ and the $(N_t + 1)$-th unit cell $(\boldsymbol{A}_{N_t+1}^a \quad \boldsymbol{B}_{N_t+1}^a)^T$ are connected through the scattering matrix $\boldsymbol{S}$

$$\begin{pmatrix} \boldsymbol{A}_{N_t+1}^{a} \\ \boldsymbol{B}_{1}^{a} \end{pmatrix} = \boldsymbol{S} \begin{pmatrix} \boldsymbol{A}_{1}^{a} \\ \boldsymbol{B}_{N_t+1}^{a} \end{pmatrix}, \tag{D19}$$

where $\boldsymbol{S}$ is computed recursively from the transfer matrix with the same procedure in Ref. [70]. Assuming the PEC BCs are set at $x = 0$ and $x = N_t\Lambda$, the explicit forms of BCs become

$$\begin{pmatrix} \cdots, & (\widetilde{\boldsymbol{E}})_{k_{+,\sigma_+}^a}, & \cdots, & (\widetilde{\boldsymbol{E}})_{k_{-,\sigma_-}^a}, & \cdots \end{pmatrix} \begin{pmatrix} \boldsymbol{A}_{1}^{a} \\ \boldsymbol{B}_{1}^{a} \end{pmatrix} = 0, \tag{D20}$$

$$\begin{pmatrix} \cdots, & (\widetilde{\boldsymbol{E}})_{k_{+,\sigma_+}^a}, & \cdots, & (\widetilde{\boldsymbol{E}})_{k_{-,\sigma_-}^a}, & \cdots \end{pmatrix} \begin{pmatrix} \boldsymbol{A}_{N_t+1}^{a} \\ \boldsymbol{B}_{N_t+1}^{a} \end{pmatrix} = 0. \tag{D21}$$

Combining Eqs. (D19–D21), the OBC spectra are then determined by the following mode condition function $g(\omega)$

$$g(\omega) = \det\left[\boldsymbol{S} + \begin{pmatrix} \mathbf{0} & \boldsymbol{I} \\ \boldsymbol{I} & \mathbf{0} \end{pmatrix}\right] = 0. \tag{D21}$$

The detailed forms of $g(\omega)$ differ under various BCs, consistent with the findings in Appendix C. However, the locations of zeros of $g(\omega)$ in the thermodynamic limit are solely determined by $\mathbf{T}(\omega)$, a point we will promptly demonstrate in Appendix E.

## Appendix E: Generalized Brillouin zone condition

The GBZ condition of 1D lattice models is commonly deduced from the Dirichlet BCs, and now we aim to generalize it to the STM-PhCs here. As mentioned in Appendix D, the eigenvalues of $\mathbf{T}(\omega)$ determine how the corresponding eigenfields evolve after translating one unit cell, so we employ $\mathbf{T}(\omega)$ to develop the non-Bloch band theory. We no longer restrict $\beta$ to the BZ in Eq. (D17), leading Eq. (D7) to the following eigenvalue equation

$$\mathbf{T}(\omega)\begin{pmatrix}\boldsymbol{A}_1^a\\ \boldsymbol{B}_1^a\end{pmatrix}_\rho = \beta_\rho\begin{pmatrix}\boldsymbol{A}_1^a\\ \boldsymbol{B}_1^a\end{pmatrix}_\rho, \tag{E1}$$

where $\beta_\rho$ and $(\boldsymbol{A}_1^a \quad \boldsymbol{B}_1^a)_\rho^T$ are the $\rho$-th eigenvalue and eigenvector of $\mathbf{T}(\omega)$, respectively. The real-space electric field corresponding to the $\rho$-th eigenvector at $x = 0$ is

$$E_{z,\rho}(0,t) = \begin{bmatrix}\sum_{\sigma_+} A_{1,\rho}^{k_{+,\sigma_+}^a}\sum_l \sqrt{\varepsilon_0}\tilde{E}_{z,l}^{k_{+,\sigma_+}^a} e^{-i(\omega+l\Omega)t}\\ +\sum_{\sigma_-} B_{1,\rho}^{k_{-,\sigma_-}^a}\sum_l \tilde{E}_{z,l}^{k_{-,\sigma_-}^a} e^{-i(\omega+l\Omega)t}\end{bmatrix}. \tag{E2}$$

Generally, the electric field at $x = 0$ can be expressed by a linear combination as

$$E_z(0,t) = \sum_\rho \begin{bmatrix}\sum_{\sigma_+} A_{1,\rho}^{k_{+,\sigma_+}^a}\sum_l \sqrt{\varepsilon_0}\tilde{E}_{z,l}^{k_{+,\sigma_+}^a} e^{-i(\omega+l\Omega)t}\\ +\sum_{\sigma_-} B_{1,\rho}^{k_{-,\sigma_-}^a}\sum_l \tilde{E}_{z,l}^{k_{-,\sigma_-}^a} e^{-i(\omega+l\Omega)t}\end{bmatrix}\phi_\rho, \tag{E3}$$

where $\phi_\rho$ represents the superposition coefficients. The according field value at $x = N_t\Lambda$ can also be easily obtained by repeatedly using Eq. (D7). We then apply the PEC BCs at $x = 0$ and $x = L = N_t\Lambda$ to the above electric fields, thus giving

$$E_z(x=0) = \sum_\rho \begin{bmatrix}\sum_{\sigma_+} A_{1,\rho}^{k_{+,\sigma_+}^a}\sum_{l=-\infty}^{+\infty} \sqrt{\varepsilon_0}\tilde{E}_{z,l}^{k_{+,\sigma_+}^a} e^{-i(\omega+l\Omega)t}\\ +\sum_{\sigma_-} B_{1,\rho}^{k_{-,\sigma_-}^a}\sum_{l=-\infty}^{+\infty} \sqrt{\varepsilon_0}\tilde{E}_{z,l}^{k_{-,\sigma_-}^a} e^{-i(\omega+l\Omega)t}\end{bmatrix}\phi_\rho = 0, \tag{E4}$$

$$E_z(x=L) = \sum_\rho \begin{bmatrix}\sum_{\sigma_+} A_{1,\rho}^{k_{+,\sigma_+}^a}\sum_{l=-\infty}^{+\infty} \sqrt{\varepsilon_0}\tilde{E}_{z,l}^{k_{+,\sigma_+}^a} e^{-i(\omega+l\Omega)t}\\ +\sum_{\sigma_-} B_{1,\rho}^{k_{-,\sigma_-}^a}\sum_{l=-\infty}^{+\infty} \sqrt{\varepsilon_0}\tilde{E}_{z,l}^{k_{-,\sigma_-}^a} e^{-i(\omega+l\Omega)t}\end{bmatrix}\beta_\rho^{N_t}\phi_\rho = 0. \tag{E5}$$

Since all the $l$ components are orthogonal, we obtain the following expression for each $l$ in Eqs. (E4-E5) as

$$\sum_\rho \left[\sum_{\sigma_+} A_{1,\rho}^{k_{+,\sigma_+}^a}\sqrt{\varepsilon_0}\tilde{E}_{z,l}^{k_{+,\sigma_+}^a} + \sum_{\sigma_-} B_{0,\rho}^{k_{-,\sigma_-}^a}\sqrt{\varepsilon_0}\tilde{E}_{z,l}^{k_{-,\sigma_-}^a}\right]\phi_\rho = 0, \tag{E6}$$

$$\sum_{\rho}\left[\sum_{\sigma_+} A_{1,\rho}^{k_{+,\sigma+}^{a}}\sqrt{\varepsilon_0}\tilde{E}_{z,l}^{k_{+,\sigma+}^{a}}+\sum_{\sigma_-} B_{0,\rho}^{k_{-,\sigma-}^{a}}\sqrt{\varepsilon_0}\tilde{E}_{z,l}^{k_{-,\sigma-}^{a}}\right]\beta_{\rho}^{N_t}\phi_{\rho}=0. \tag{E7}$$

As the value of $l$ ranges from $-l_c$ to $+l_c$, Eqs. (E6) and (E7) yield $2l_c+1$ equations each. The condition for the equation set to have non-zero solutions is that the determinant of the coefficient matrix is equal to zero, which can be explicitly written as

$$\begin{vmatrix} g_{-l_c,1} & g_{-l_c,2} & \cdots & g_{-l_c,2*(2*l_c+1)} \\ g_{-l_c+1,1} & g_{-l_c+1,2} & \cdots & g_{-l_c+1,2*(2*l_c+1)} \\ \vdots & \vdots & \ddots & \vdots \\ g_{l_c,1} & g_{l_c,2} & \cdots & g_{l_c,2*(2*l_c+1)} \\ g_{-l_c,1}\beta_1^{N_t} & g_{-l_c,2}\beta_2^{N_t} & \cdots & g_{-l_c,2*(2*l_c+1)}\beta_{2*(2*l_c+1)}^{N_t} \\ g_{-l_c+1,1}\beta_1^{N_t} & g_{-l_c+1,2}\beta_2^{N_t} & \cdots & g_{-l_c+1,2*(2*l_c+1)}\beta_{2*(2*l_c+1)}^{N_t} \\ \vdots & \vdots & \ddots & \vdots \\ g_{l_c,1}\beta_1^{N_t} & g_{l_c,2}\beta_2^{N_t} & \cdots & g_{l_c,2*(2*l_c+1)}\beta_{2*(2*l_c+1)}^{N_t} \end{vmatrix}=0, \tag{E8}$$

where $g_{l,\rho}=\left[\sum_{\sigma_+} A_{1,\rho}^{k_{\sigma+}^{a}}\sqrt{\varepsilon_0}\tilde{E}_{z,l}^{k_{+,\sigma+}^{a}}+\sum_{\sigma_-} B_{1,\rho}^{k_{\sigma-}^{a}}\sqrt{\varepsilon_0}\tilde{E}_{z,l}^{k_{-,\sigma-}^{a}}\right]$ and $\beta$ are sorted by their magnitudes, satisfying $|\beta_1|\le|\beta_2|\le\cdots\le|\beta_{2(2l_c+1)-1}|\le|\beta_{2(2l_c+1)}|$. Expand Eq. (E8) and explicitly write out two leading terms as

$$C_1\beta_{(2l_c+2)}^{N_t}\beta_{(2l_c+3)}^{N_t}\cdots\beta_{2(2l_c+1)}^{N_t}+C_2\beta_{(2l_c+1)}^{N_t}\beta_{(2l_c+3)}^{N_t}\cdots\beta_{2(2l_c+1)}^{N_t}+\text{other terms}=0. \tag{E9}$$

Here, we have arbitrarily used the symbols $C_1$ and $C_2$ to represent the coefficients of two leading terms. Importantly, they are independent of the system size. As $L$ approaches infinity, based on the consideration of the continuum spectra [31], we can obtain

$$|\beta_{2l_c+1}|=|\beta_{2l_c+2}|. \tag{E10}$$

Since $\beta$ represents the eigenvalues of $\mathbf{T}(\omega)$, we can directly use the $\mathbf{T}(\omega)$ matrix of a single unit cell to solve the OBC spectrum in the thermodynamic limit by using Eqs. (E10). The corresponding $\beta$ set constitutes the GBZ. Note that the truncation $l_c$ in the PWE may bring numerical errors. Hence, a preferable approach is to initially use a large value of $l_c$ to construct the $\mathbf{T}(\omega)$ matrix and subsequently extract a smaller but numerically stable $\mathbf{T}(\omega)$ to mitigate the numerical issues arising from the finite truncation.

## Appendix F: Calculation of biorthogonal Berry connection

In non-Hermitian systems, the conventional inner product shall be generalized to the biorthogonal form, thus also necessitating the biorthogonal Berry connection. However, acquiring the left eigenstates is not that straightforward, especially for a system without an explicit Hamiltonian matrix, which is precisely the scenario here. Technically speaking, we need to find the adjoint of the original system. To be more generic, we consider the three-dimensional Maxwell equations by redefining the quantities in Eq. (20) as

$$\hat{\boldsymbol{L}}=\begin{pmatrix}0 & i\nabla\times\\ -i\nabla\times & 0\end{pmatrix},\boldsymbol{K}=\begin{pmatrix}\boldsymbol{\varepsilon} & \boldsymbol{\xi}\\ \boldsymbol{\eta} & \boldsymbol{\mu}\end{pmatrix},\boldsymbol{\psi}^R=\begin{pmatrix}\sqrt{\varepsilon_0}\boldsymbol{E}\\ \sqrt{\mu_0}\boldsymbol{H}\end{pmatrix}, \tag{F1}$$

where the argument $x$ and the subscript $E$ that represents the E-PhC have been omitted for simplicity, and the superscript $R$ denotes the right eigenvector. The generalized eigenvalue equation (19) is then reformulated as

$$\hat{\boldsymbol{L}}_K\boldsymbol{\psi}^R=\frac{\omega}{c}\boldsymbol{\psi}^R,\ \hat{\boldsymbol{L}}_K=\boldsymbol{K}^{-1}\hat{\boldsymbol{L}}. \tag{F2}$$

By defining the inner product as $\langle\boldsymbol{\psi}_1,\boldsymbol{\psi}_2\rangle=\int \mathrm{d}\boldsymbol{r}\ \boldsymbol{\psi}_1^\dagger\cdot\boldsymbol{\psi}_2$ and with the help of an identity $\nabla\cdot(\boldsymbol{f}\times\boldsymbol{g})=(\nabla\times\boldsymbol{f})\cdot\boldsymbol{g}-\boldsymbol{f}\cdot(\nabla\times\boldsymbol{g})$, we can obtain the formal adjoint operator $\bar{\boldsymbol{L}}_K$ of $\hat{\boldsymbol{L}}_K$ as

$$\langle\boldsymbol{\psi}_1,\hat{\boldsymbol{L}}_K\boldsymbol{\psi}_2\rangle=\langle\bar{\boldsymbol{L}}_K\boldsymbol{\psi}_1,\boldsymbol{\psi}_2\rangle+J(\boldsymbol{\psi}_1,\boldsymbol{\psi}_2),\qquad \bar{\boldsymbol{L}}_K=\hat{\boldsymbol{L}}(\boldsymbol{K}^{-1})^\dagger, \tag{F3}$$

where $J(\boldsymbol{\psi}_1,\boldsymbol{\psi}_2)$ is the bilinear concomitant or conjunct, coming from the BCs [83]. The explicit form of $J(\boldsymbol{\psi}_1,\boldsymbol{\psi}_2)$ has not been put down here, but specifically for PBCs used in the Berry connection, $J(\boldsymbol{\psi}_1,\boldsymbol{\psi}_2)$ is zero. Therefore, $\bar{\boldsymbol{L}}_K$ is indeed the adjoint operator of $\hat{\boldsymbol{L}}_K$. The eigenvalue problem of $\bar{\boldsymbol{L}}_K$ with the eigenvalue and the left eigenvector herein is defined as

$$\bar{\boldsymbol{L}}_K\boldsymbol{\psi}_{m_i}^L=\frac{\omega_{m_i}^*}{c}\boldsymbol{\psi}_{m_i}^L, \tag{F4}$$

where $m_i$ is now the eigenvalue index, later becoming the band index. Together with Eqs. (F2-F3), we arrive at the following biorthogonal relation

$$\left(\frac{\omega_{n_i}}{c}-\frac{\omega_{m_i}}{c}\right)\langle\boldsymbol{\psi}_{m_i}^L,\boldsymbol{\psi}_{n_i}^R\rangle=0, \tag{F5}$$

where $n_i$ is also the eigenvalue index. When $m_i=n_i$ ($m_i\neq n_i$), $\langle\boldsymbol{\psi}_{m_i}^L,\boldsymbol{\psi}_{n_i}^R\rangle\neq 0$ ($\langle\boldsymbol{\psi}_{m_i}^L,\boldsymbol{\psi}_{n_i}^R\rangle=0$) gives the normalization condition (biorthogonal relation). The central stuff now is to relate the adjoint eigenvalue problem with the original one.

By taking the complex conjugate of Eq. (F4)

$$\bar{\boldsymbol{L}}_K^*\boldsymbol{\psi}_{m_i}^{L,*}=\frac{\omega_{m_i}}{c}\boldsymbol{\psi}_{m_i}^{L,*}, \tag{F6}$$

and using Eqs. (F1-F3), we can obtain

$$-\hat{\boldsymbol{L}}_K(\boldsymbol{K}^{-1})^{\mathrm{T}}\boldsymbol{\psi}_{m_i}^{L,*}=\frac{\omega_{m_i}}{c}\boldsymbol{K}^{-1}\boldsymbol{\psi}_{m_i}^{L,*}. \tag{F7}$$

If $(\boldsymbol{K}^{-1})^{\mathrm{T}}=\boldsymbol{K}^{-1}$, which is apparent in the E-PhC [see Eqs. (4) and (20)], then $\boldsymbol{K}^{-1}\boldsymbol{\psi}_{m_i}^{L,*}$ is an eigenstate of the $\hat{\boldsymbol{L}}_K$ operator, with the corresponding eigenvalue being $-\omega_{m_i}/c$. This indicates that in our scenario, we can use the right eigenstate of $\hat{\boldsymbol{L}}_K$ at the eigenvalue $-\omega_{m_i}/c$ to calculate the left eigenstate $\boldsymbol{\psi}_{m_i}^L$. Note that $(\boldsymbol{K}^{-1})^{\mathrm{T}} = \boldsymbol{K}^{-1}$ is not the Lorentz reciprocal condition [41].

Concerning the PhC, the periodicity in $\hat{\boldsymbol{L}}_K$ implies $\hat{T}_\Lambda \boldsymbol{\psi}_{n_i,q}^R = e^{iq\Lambda}\boldsymbol{\psi}_{n_i,q}^R$, where $\hat{T}_\Lambda$ is the translational operator and $q$ is the Bloch wavenumber. The biorthogonal normalization relation $\langle \boldsymbol{\psi}_{n_i,q}^L, \boldsymbol{\psi}_{n_i,q}^R \rangle = 1$ further indicates $\hat{T}_\Lambda \boldsymbol{\psi}_{n_i,q}^L = e^{iq\Lambda}\boldsymbol{\psi}_{n_i,q}^L$, and thus, we have

$$\hat{T}_\Lambda \boldsymbol{K}^{-1}\boldsymbol{\psi}_{n_i,q}^{L,*} = e^{-iq\Lambda}\boldsymbol{K}^{-1}\boldsymbol{\psi}_{n_i,q}^{L,*}, \tag{F8}$$

which shows that $\boldsymbol{K}^{-1}\boldsymbol{\psi}_{n_i,q}^{L,*}$ is an eigenstate at $-q$. Equations (F7-F8) together declare the relationship between the left and right eigenstates of the $\hat{\boldsymbol{L}}_{\boldsymbol{K}}$ operator

$$\boldsymbol{K}^{-1}\boldsymbol{\psi}_{n_i,q}^{L,*} = \boldsymbol{\psi}_{-n_i,-q}^R, \tag{F9}$$

where $-n_i$ represents the energy bands at negative frequencies $-\omega_{n_i}(-q)/c$. The corresponding biorthogonal normalization relation is then rewritten as

$$\delta_{m_i,n_i} = \int_{\mathrm{uc}} \mathrm{d}x \left(\boldsymbol{\psi}_{m_i,q}^L\right)^\dagger \boldsymbol{\psi}_{n_i,q}^R = \int_{\mathrm{uc}} \mathrm{d}x \left(\boldsymbol{u}_{-m_i,-q}^R\right)^{\mathrm{T}} \boldsymbol{K}\boldsymbol{u}_{n_i,q}^R, \tag{F10}$$

where $\boldsymbol{u}_{n_i,q}^R$ is the periodic part of $\boldsymbol{\psi}_{n_i,q}^R = \boldsymbol{u}_{n_i,q}^R e^{iq\Lambda}$. From Eq. (F10), we generalize the Berry connection for isolated bands as

$$\boldsymbol{A}_{n_i}(q) = i\langle u_{n_i,q}^L \partial_q | u_{n_i,q}^R \rangle = i\int_{\mathrm{uc}} \mathrm{d}x \left(\boldsymbol{u}_{-n_i,-q}^R\right)^{\mathrm{T}} \boldsymbol{K}\partial_q \boldsymbol{u}_{n_i,q}^R. \tag{F11}$$

In the non-Hermitian topology, the integral for topological invariants should be performed on the GBZ as

$$\theta_{n_i}^{\mathrm{Zak}} = \oint_{\mathrm{GBZ}_{\mathrm{n_i}}} d\varphi_\beta \left[ i\left\langle u_{n_i,\varphi_\beta}^L \middle| \partial_{\varphi_\beta} u_{n_i,\varphi_\beta}^R \right\rangle \right], \tag{F12}$$

where $\varphi_\beta$ represents the phase angle of $\beta$ along the GBZ. Because of the relation Eq. (F9), the left state $u_{n_i,\varphi_\beta}^L$ shall take the state residing on the adjoint GBZ, as shown by solid gray lines in Fig. 4.

**References**

[1] L. Feng, R. El-Ganainy, and L. Ge, Nat. Photon. **11**, 752 (2017).

[2] R. El-Ganainy, K. G. Makris, M. Khajavikhan, Z. H. Musslimani, S. Rotter, and D. N. Christodoulides, Nat. Phys. **14**, 11 (2018).

[3] Y. Ashida, Z. Gong, and M. Ueda, Adv. Phys. **69**, 249 (2020).

[4] E. J. Bergholtz, J. C. Budich, and F. K. Kunst, Rev. Mod. Phys. **93**, 015005 (2021).

[5] K. Ding, C. Fang, and G. Ma, Nat. Rev. Phys. **4**, 745 (2022).

[6] N. Okuma and M. Sato, Annu. Rev. Condens. Matter Phys. **14**, 83 (2023).

[7] C. M. Bender and S. Boettcher, Phys. Rev. Lett. **80**, 5243 (1998).

[8] C. M. Bender, D. C. Brody, and H. F. Jones, Phys. Rev. Lett. **89**, 270401 (2002).

[9] M. B. Carl, M. V. Berry, and M. Aikaterini, J. Phys. A: Math. Gen. **35**, L467 (2002).

[10] A. Mostafazadeh, J. Math. Phys. **43**, 205 (2002).

[11] A. Mostafazadeh, J. Math. Phys. **43**, 2814 (2002).

[12] A. Mostafazadeh, J. Math. Phys. **43**, 3944 (2002).

[13] K. Ding, Z. Q. Zhang, and C. T. Chan, Phys. Rev. B **92**, 235310 (2015).

[14] H. Shen, B. Zhen, and L. Fu, Phys. Rev. Lett. **120**, 146402 (2018).

[15] X. Cui, K. Ding, J.-W. Dong, and C. T. Chan, Phys. Rev. B **100**, 115412 (2019).

[16] K. Kawabata, K. Shiozaki, M. Ueda, and M. Sato, Phys. Rev. X **9**, 041015 (2019).

[17] J. Hu, R.-Y. Zhang, Y. Wang, X. Ouyang, Y. Zhu, H. Jia, and C. T. Chan, Nat. Phys. **19**, 1098 (2023).

[18] W. Tang, K. Ding, and G. Ma, Nat. Commun. **14**, 6660 (2023).

[19] K. Wang, A. Dutt, C. C. Wojcik, and S. Fan, Nature **598**, 59 (2021).

[20] Y. S. S. Patil, J. Höller, P. A. Henry, C. Guria, Y. Zhang, L. Jiang, N. Kralj, N. Read, and J. G. E. Harris, Nature **607**, 271 (2022).

[21] C.-X. Guo, S. Chen, K. Ding, and H. Hu, Phys. Rev. Lett. **130**, 157201 (2023).

[22] K. Ding, G. Ma, M. Xiao, Z. Q. Zhang, and C. T. Chan, Phys. Rev. X **6**, 021007 (2016).

[23] W. Chen, Ş. Kaya Özdemir, G. Zhao, J. Wiersig, and L. Yang, Nature **548**, 192 (2017).

[24] K. Ding, G. Ma, Z. Q. Zhang, and C. T. Chan, Phys. Rev. Lett. **121**, 085702 (2018).

[25] W. Tang, X. Jiang, K. Ding, Y.-X. Xiao, Z.-Q. Zhang, C. T. Chan, and G. Ma, Science **370**, 1077 (2020).

[26] S. Yao and Z. Wang, Phys. Rev. Lett. **121**, 086803 (2018).

[27] F. K. Kunst, E. Edvardsson, J. C. Budich, and E. J. Bergholtz, Phys. Rev. Lett. **121**, 026808 (2018).

[28] T. Helbig *et al.*, Nat. Phys. **16**, 747 (2020).

[29] K. Zhang, Z. Yang, and C. Fang, Phys. Rev. Lett. **125**, 126402 (2020).
[30] N. Okuma, K. Kawabata, K. Shiozaki, and M. Sato, Phys. Rev. Lett. **124**, 086801 (2020).
[31] K. Yokomizo and S. Murakami, Phys. Rev. Lett. **123**, 066404 (2019).
[32] Z. Yang, K. Zhang, C. Fang, and J. Hu, Phys. Rev. Lett. **125**, 226402 (2020).
[33] Y.-M. Hu, Y.-Q. Huang, W.-T. Xue, and Z. Wang, arXiv:2310.08572 (2023).
[34] K. Zhang, Z. Yang, and C. Fang, Nat. Commun. **13**, 2496 (2022).
[35] H.-Y. Wang, F. Song, and Z. Wang, arXiv:2212.11743 (2022).
[36] K. Yokomizo and S. Murakami, Phys. Rev. B **107**, 195112 (2023).
[37] K. Zhang, C. Fang, and Z. Yang, Phys. Rev. Lett. **131**, 036402 (2023).
[38] K. Zhang, Z. Yang, and K. Sun, arXiv:2309.03950 (2023).
[39] Z. Xu, B. Pang, K. Zhang, and Z. Yang, arXiv:2311.16868 (2023).
[40] W. Wang, M. Hu, X. Wang, G. Ma, and K. Ding, Phys. Rev. Lett. **131**, 207201 (2023).
[41] J. A. Kong, *Electromagnetic Wave Theory* (Wiley-Interscience, 1986).
[42] C. Caloz, A. Alù, S. Tretyakov, D. Sounas, K. Achouri, and Z.-L. Deck-Léger, Phys. Rev. Appl. **10**, 047001 (2018).
[43] R. Tirole, S. Vezzoli, E. Galiffi, I. Robertson, D. Maurice, B. Tilmann, S. A. Maier, J. B. Pendry, and R. Sapienza, Nat. Phys. **19**, 999 (2023).
[44] P. A. Huidobro, E. Galiffi, S. Guenneau, R. V. Craster, and J. B. Pendry, Proc. Natl. Acad. Sci. U. S. A. **116**, 24943 (2019).
[45] F. R. Prudêncio and M. G. Silveirinha, Phys. Rev. Appl. **19**, 024031 (2023).
[46] E. Lustig, Y. Sharabi, and M. Segev, Optica **5**, 1390 (2018).
[47] M. Lyubarov, Y. Lumer, A. Dikopoltsev, E. Lustig, Y. Sharabi, and M. Segev, Science **377**, 425 (2022).
[48] P. Kongkhambut, J. Skulte, L. Mathey, J. G. Cosme, A. Hemmerich, and H. Keßler, Science **377**, 670 (2022).
[49] C. Caloz and Z. L. Deck-Léger, IEEE Trans. Antennas Propag. **68**, 1569 (2020).
[50] V. Pacheco-Peña and N. Engheta, Optica **7**, 323 (2020).
[51] V. Pacheco-Peña and N. Engheta, Phys. Rev. B **104**, 214308 (2021).
[52] C. Guo, M. Xiao, M. Orenstein, and S. Fan, Light Sci. Appl. **10**, 160 (2021).
[53] G.-B. Wu, J. Y. Dai, Q. Cheng, T. J. Cui, and C. H. Chan, Nat. Electron. **5**, 808 (2022).
[54] F. R. Prudêncio and M. G. Silveirinha, Nanophotonics **12**, 3007 (2023).
[55] N. Chamanara, Z.-L. Deck-Léger, C. Caloz, and D. Kalluri, Phys. Rev. A **97**, 063829 (2018).
[56] N. Wang, Z.-Q. Zhang, and C. T. Chan, Phys. Rev. B **98**, 085142 (2018).

[57] Y. Sharabi, A. Dikopoltsev, E. Lustig, Y. Lumer, and M. Segev, Optica **9**, 585 (2022).
[58] M. Moghaddaszadeh, M. A. Attarzadeh, A. Aref, and M. Nouh, Phys. Rev. Appl. **18**, 044013 (2022).
[59] K. Yokomizo, T. Yoda, and S. Murakami, Phys. Rev. Res. **4**, 023089 (2022).
[60] T. Yoda, Y. Moritake, K. Takata, K. Yokomizo, S. Murakami, and M. Notomi, arXiv:2303.05185 (2023).
[61] M. Davanco, Y. Urzhumov, and G. Shvets, Opt. Express **15**, 9681 (2007).
[62] C. Fietz, Y. Urzhumov, and G. Shvets, Opt. Express **19**, 19027 (2011).
[63] J. H. D. Rivero, L. Feng, and L. Ge, Phys. Rev. Lett. **129**, 243901 (2022).
[64] A. Yariv and P. A. Yeh, *Optical Waves in Crystals: Propagation and Control of Laser Radiation* (Wiley, New York, 1983).
[65] P. A. Huidobro, M. G. Silveirinha, E. Galiffi, and J. B. Pendry, Phys. Rev. Appl. **16**, 014044 (2021).
[66] Q. Yan, H. Chen, and Y. Yang, Prog. Electromagn. Res. **172**, 33 (2021).
[67] J. B. Pendry, E. Galiffi, and P. A. Huidobro, J. Opt. Soc. Am. B **38**, 3360 (2021).
[68] J. B. Pendry, E. Galiffi, and P. A. Huidobro, Optica **9**, 724 (2022).
[69] E. Galiffi, P. A. Huidobro, and J. B. Pendry, Phys. Rev. Lett. **123**, 206101 (2019).
[70] L. Li, J. Opt. Soc. Am. A **13**, 1024 (1996).
[71] K. Sakoda, *Optical Properties of Photonic Crystals* (Springer Berlin, Heidelberg, 2004).
[72] K. Yokomizo, T. Yoda, and Y. Ashida, arXiv:2311.15553 (2023).
[73] J. Zak, Phys. Rev. Lett. **62**, 2747 (1989).
[74] M. Xiao, Z. Q. Zhang, and C. T. Chan, Physical Review X **4**, 021017 (2014).
[75] D. Vanderbilt, *Berry Phases in Electronic Structure Theory: Electric Polarization, Orbital Magnetization and Topological Insulators* (Cambridge University Press, Cambridge, 2018).
[76] W. Zhu, W. X. Teo, L. Li, and J. Gong, Phys. Rev. B **103**, 195414 (2021).
[77] W. Wang, X. Wang, and G. Ma, Nature **608**, 50 (2022).
[78] W. Wang, X. Wang, and G. Ma, Phys. Rev. Lett. **129**, 264301 (2022).
[79] T. Li and H. Hu, Nat. Commun. **14**, 6418 (2023).
[80] R. Slager, A. Bouhon, and F. Ünal, arXiv:2310.12782 (2023).
[81] S. Xu and C. Wu, Phys. Rev. Lett. **120**, 096401 (2018).
[82] Q. Gao and Q. Niu, Phys. Rev. Lett. **127**, 036401 (2021).

[83] G. B. Arfken, H. J. Weber, and F. E. Harris, in *Mathematical Methods for Physicists (Seventh Edition)*, edited by G. B. Arfken, H. J. Weber, and F. E. Harris (Academic Press, Boston, 2013), pp. 381.

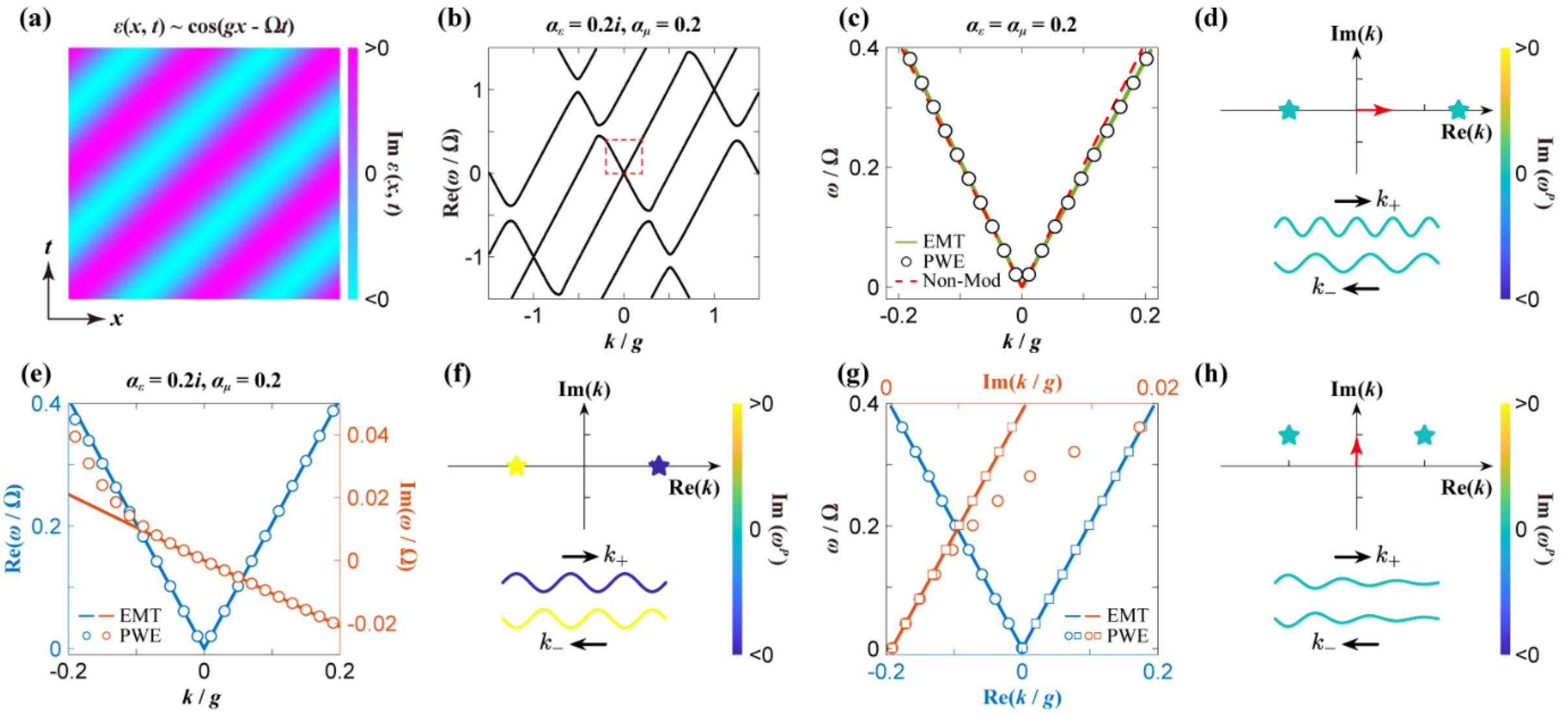

FIG.1 (a) Schematic of the permittivity distribution in the space-time domain of an STM. (b) A typical Floquet BS for the STM, wherein only real parts of frequency are plotted. (c) The PBC bands in the long-wavelength limit [the region highlighted by the dashed box in (b)] with purely real modulation $\alpha_\varepsilon = \alpha_\mu = 0.2$. The red dashed lines are the bands without modulations ($\alpha_\varepsilon = \alpha_\mu = 0$). (e) The PBC BS and (g) complex-$k$ BS of an STM with $\alpha_\varepsilon = 0.2i$ and $\alpha_\mu = 0.2$. The solid lines and open markers in (c,e,g) represent the EMT and PWE results, respectively. Only one solid red line is seen in (g) because two $\mathrm{Im}(k)$ branches are identical. (d, f, h) depict the diagrammatical understanding of the results in (c, e, g). The top panels of (d, f) depict the EFC of a representative real frequency with the color of the markers showing $\mathrm{Im}(\omega^P)$. The top panel of (h) shows the $k$ position for a representative real frequency in the complex-$k$ BS. The bottom panels sketch the wave propagation features correspondingly. Other system parameters are $\varepsilon_r = 6$, $\mu_r = 1$, $\phi_\varepsilon = \phi_\mu = 0$, and $\Omega/g = 0.2c$. The cutoff in the PWE is set to $l_c = 15$.

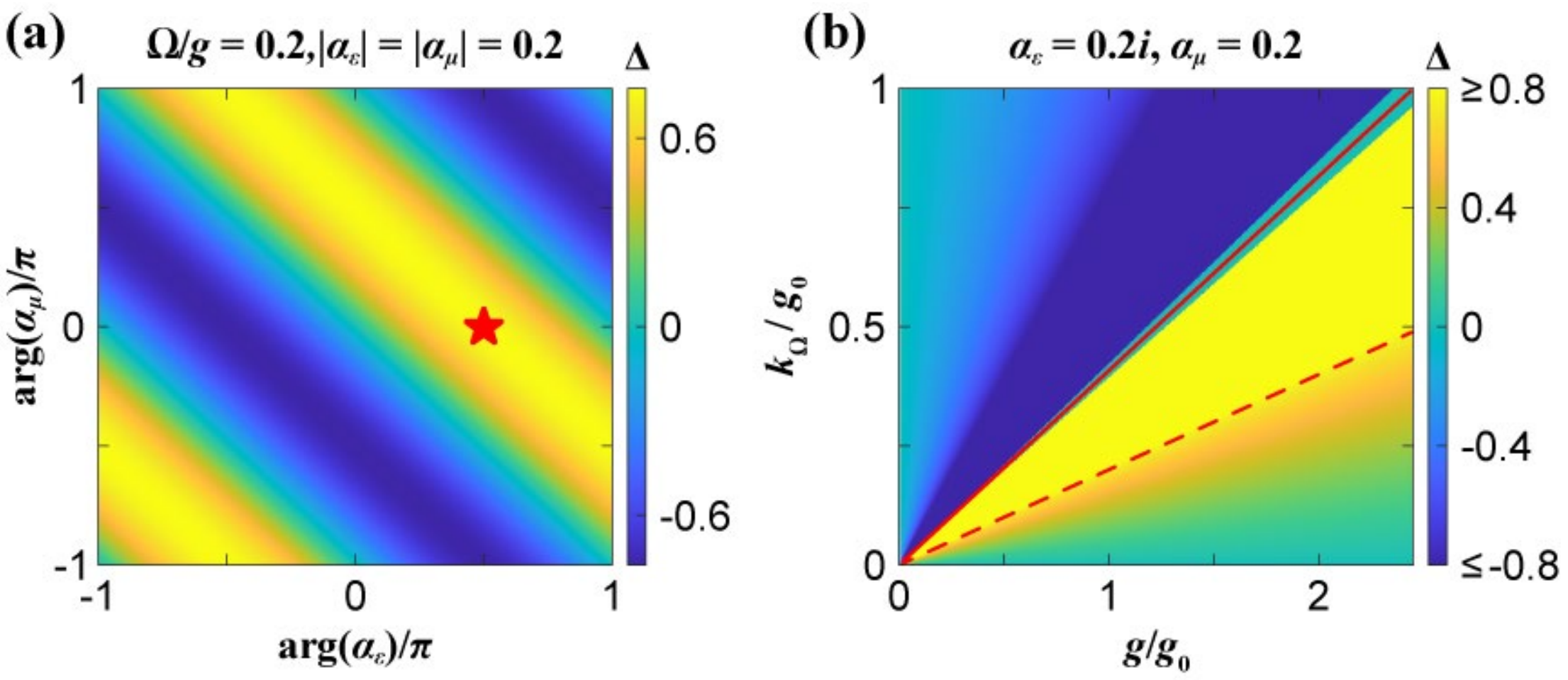

FIG.2 Contour plot of $\Delta$ in the $\arg(\alpha_\varepsilon)$-$\arg(\alpha_\mu)$ plane (a) and the $g$-$k_\Omega$ plane (b). The red pentagram in (a) denotes the scenario of Figs. 1(e) and 1(g). The red solid line (dashed line) in (b) represents the speed of light in the background material (the modulation speed in Fig. 1). $g_0$ is a parameter that accounts for the concrete physical system, facilitating the nondimensional coordinates here. The background permittivity and permeability are set to $\varepsilon_r = 6$ and $\mu_r = 1$. All other parameters are indicated in the figures.

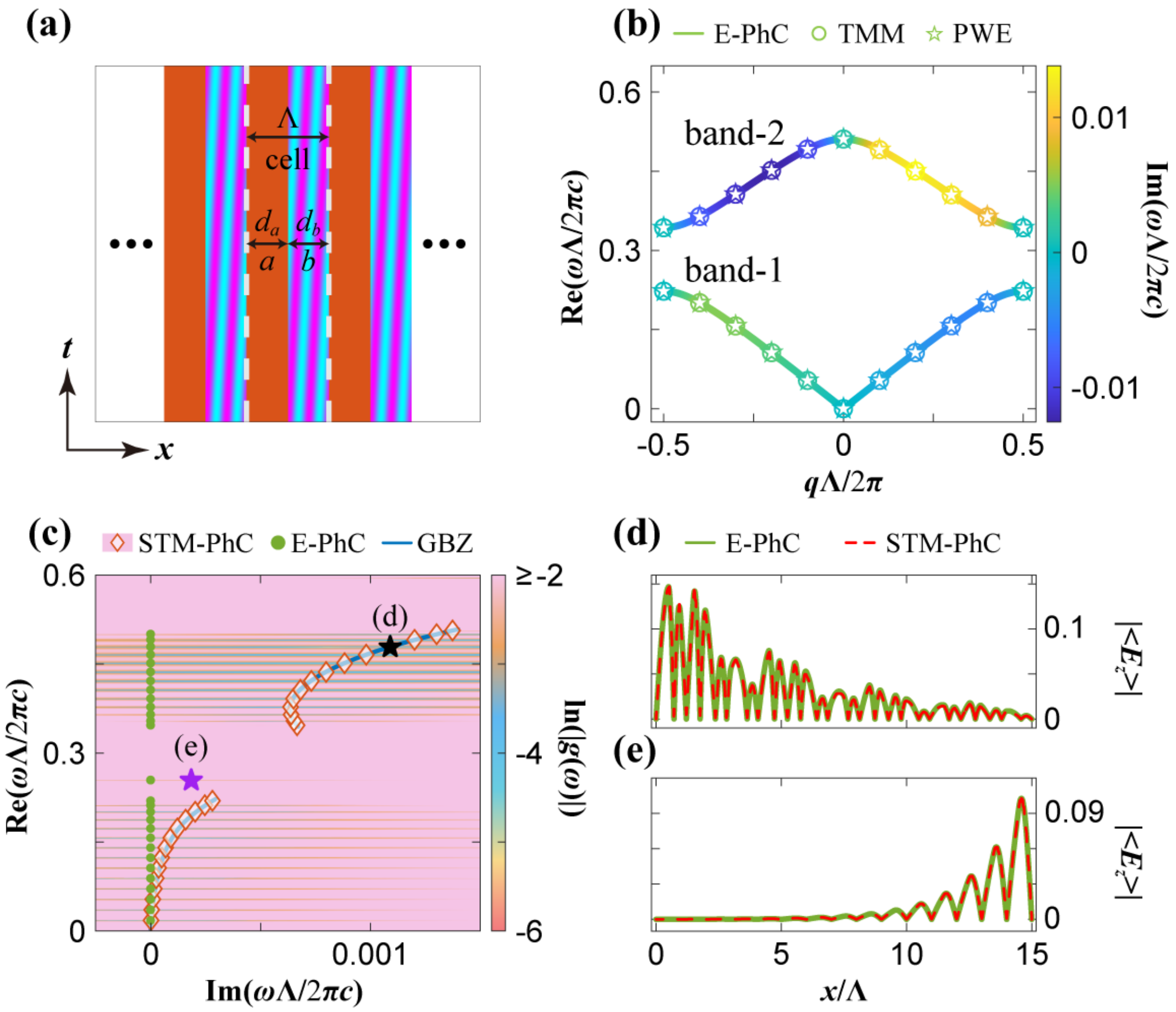

FIG.3 (a) Schematic of a spatial PhC composed of homogeneous materials (component $a$) and STMs (component $b$). The unit cell is labeled herein with its lattice constant $\Lambda = 2\pi N/g$. (b) The PBC BS of an STM-PhC with its unit cell shown in (a). The color of all these plots represents $\mathrm{Im}(\omega^P)$. Solid lines, open circles, and open stars correspond to numerical results of the E-PhC, TMM, and PWE, respectively. The frequency $\mathrm{Re}(\omega\Lambda/2\pi c) = 0.6$ here corresponds to $\mathrm{Re}(\omega/\Omega) = 0.15$ in Fig. 1. (c) The spectrum of a finite-sized STM-PhC under PEC BCs. The color plot shows $|g(\omega)|$ with its zeros marked by the red diamonds. The green dots and blue lines represent the results from E-PhC and GBZ, respectively. The field distributions of a bulk state (black pentagram) and the topological edge state (purple pentagram) are shown in (d) and (e), respectively. The green (red) lines are from the E-PhC (STM-PhC). The parameters of the component $a$ are $\varepsilon_r = \mu_r = 1$ and $d_a/\Lambda = 0.5$ ($N = 20$), while those of the component $b$ are the same as Figs. 1(e) and 1(g). The number of unit cells is $N_t = 15$. The cutoff is set to $l_c = 2$, $n_c = 60$, and $o_c = 100$.

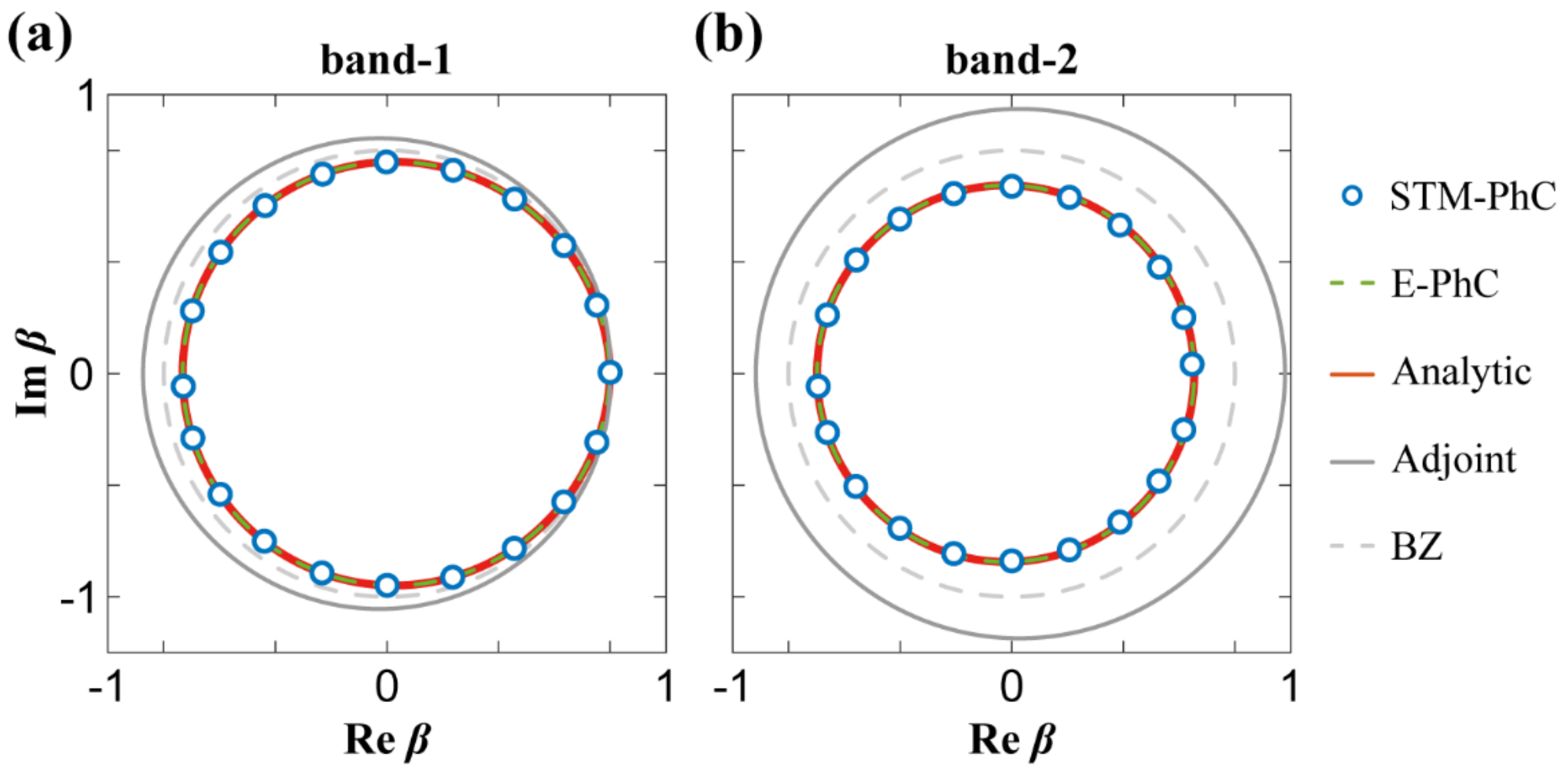

FIG.4 The GBZs for band-1 and band-2 in Fig. 3(b) are depicted in (a) and (b), respectively. The blue markers, green dashed lines, and red solid lines represent the GBZs calculated using the STM-PhC, the E-PhC, and analytical methods. The solid gray line represents the adjoint GBZ, and the gray dashed line depicts the BZ.

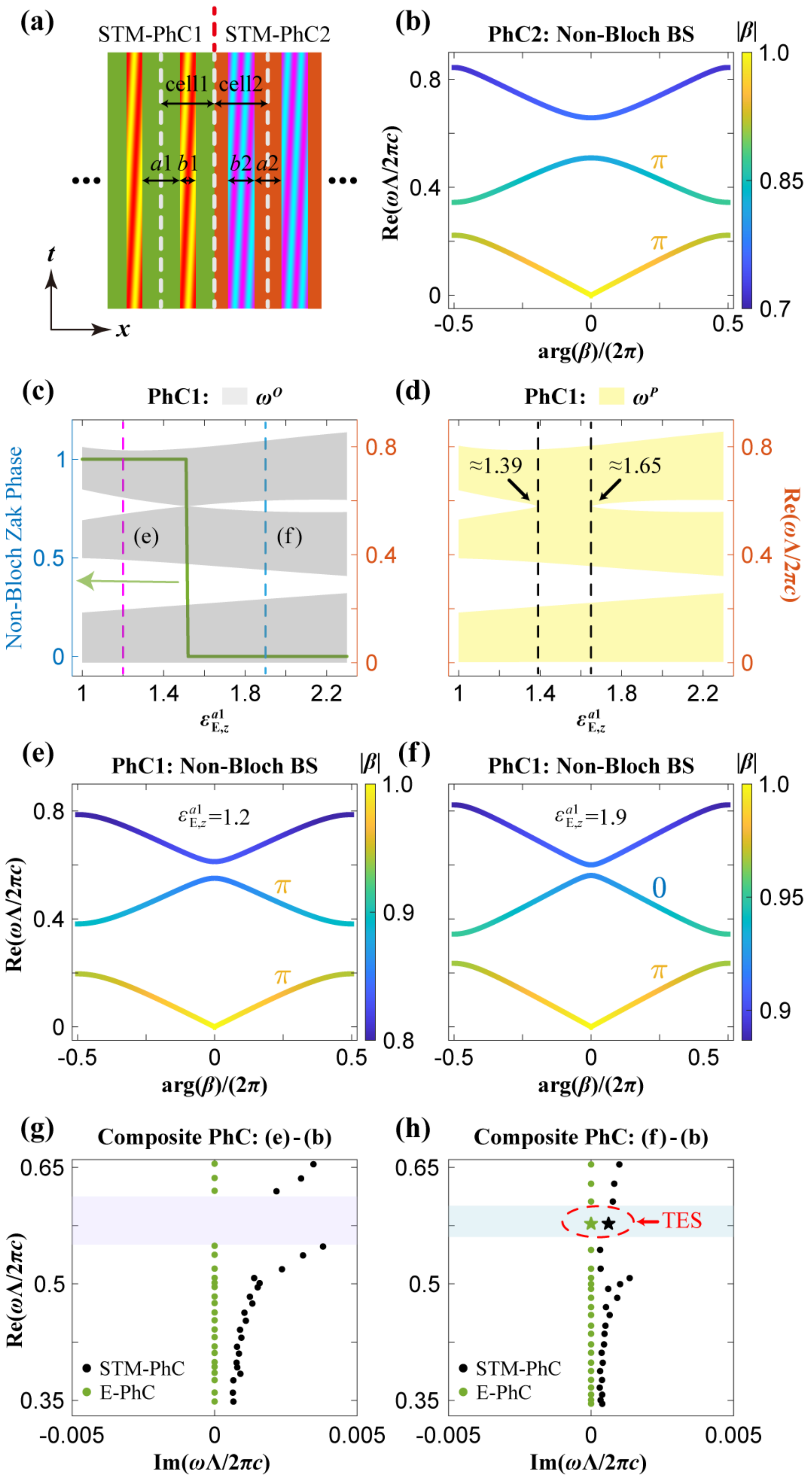

(a)
STM-PhC1
STM-PhC2
cell1
cell2
a1 b1
b2 a2
t
x
(b)
PhC2: Non-Bloch BS
|β|
Re(ωΛ/2πc)
arg(β)/(2π)
π
π
(c)
PhC1: ω^O
Non-Bloch Zak Phase
(e)
(f)
(d)
PhC1: ω^P
≈1.39
≈1.65
Re(ωΛ/2πc)
(e)
PhC1: Non-Bloch BS
ε^{a1}_{E,z}=1.2
(f)
PhC1: Non-Bloch BS
ε^{a1}_{E,z}=1.9
0
(g)
Composite PhC: (e) - (b)
STM-PhC
E-PhC
Im(ωΛ/2πc)
(h)
Composite PhC: (f) - (b)
TES

FIG.5 (a) Schematics of a domain wall formed by two STM-PhCs. (b) The non-Bloch BS $\mathrm{Re}(\omega\Lambda/2\pi c)$-$\arg(\beta)/2\pi$ of the STM-PhC2 with the line color representing $|\beta|$. The non-Bloch Zak phase of each non-Bloch band is indicated nearby. The parameters herein are the same as in Fig. 3, except for the symmetric unit cell adopted here. (c) The non-Bloch Zak phase (left y-axis) and the OBC spectra (right y-axis) of the STM-PhC1 as a function of $\varepsilon_{\mathrm{E,z}}^{a1}$. (d) The PBC spectra of the STM-PhC1 as a function of $\varepsilon_{\mathrm{E,z}}^{a1}$. The parameters of the STM-PhC1 fixed in (c-d) are $\mu_{\mathrm{E,y}}^{a1} = \mu_{\mathrm{E,y}}^{b1} = 1$, $d_{a1}/\Lambda = 0.7$, $\alpha_\varepsilon = 0.15i$, and $\alpha_\mu = 0.15$. We adjust the values of $\varepsilon_{\mathrm{E,z}}^{b1}$ accordingly with $\varepsilon_{\mathrm{E,z}}^{a1}$ to align the second non-Bloch band gap. (e,f) The non-Bloch BS with the corresponding non-Bloch Zak phase labeled of the STM-PhC1 when (e) $\varepsilon_{\mathrm{E,z}}^{a1} = 1.2$ [the magenta dashed line in (c)] and (f) $\varepsilon_{\mathrm{E,z}}^{a1} = 1.9$ [the blue dashed line in (c)]. The numerical calculations in (b-f) are performed based on the E-PhC corresponding to the STM-PhC. (g,h) The OBC spectra of the composite PhC composed of the PhCs defined in (e) and (b) [(f) and (b)] are shown in (g) [(h)]. The green (black) markers represent the results obtained using the E-PhC (STM-PhC) setup, in which the circles (pentagrams) denote the bulk states (TESs). The purple and blue shaded regions highlight common non-Bloch band gaps of the two PhCs. The number of unit cells for both PhCs in the OBC calculations is chosen to be ten.

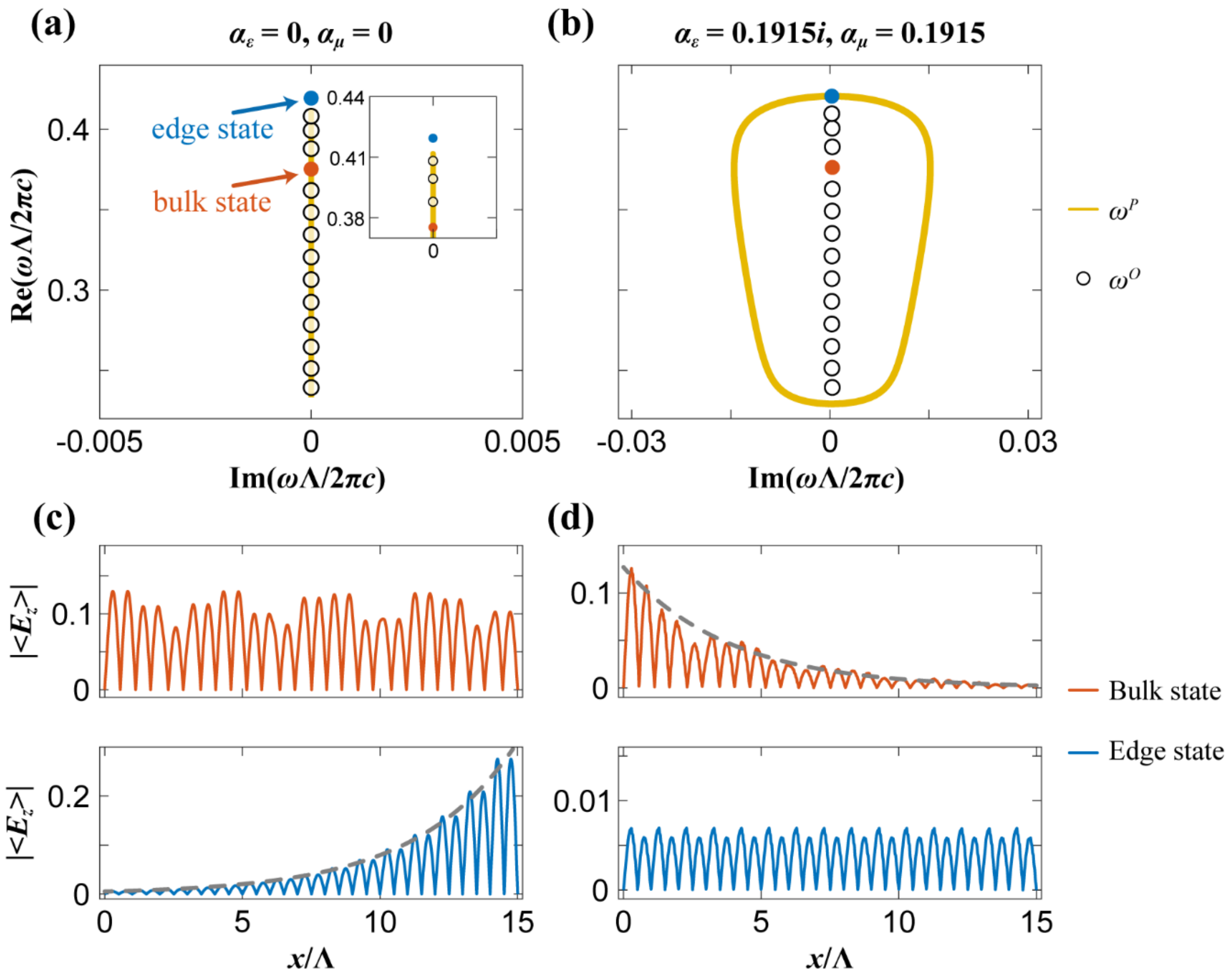

FIG.6 (a,b) The OBC spectra (circle markers) and PBC spectra (yellow lines) for band-2 in Fig. 3(a) when (a) $\alpha_\varepsilon = \alpha_\mu = 0$ and (b) $\alpha_\varepsilon = 0.1915i, \alpha_\mu = 0.1915$. The inset in (a) is a magnification of the spectra near the gap. The filled blue and red circles, respectively, denote the TES and one representative bulk state, with their field distributions depicted in (c) and (d). The gray dashed lines in the bottom panel of (c) [top panel of (d)] exponentially fit the envelope of localized states by $|\langle E_z \rangle| \propto \exp(\kappa_{\text{fit}}\, x/\Lambda)$ with $\kappa_{\text{fit}} = +0.27$ [$\kappa_{\text{fit}} = -0.26$]. Except for $d_a = 0.15\Lambda$ and $d_b = 0.85\Lambda$, other parameters are the same as Fig. 3.